\documentclass[conference]{IEEEtran}
\usepackage{cite}
\usepackage{dsfont}
\usepackage{amsmath,amsthm,amssymb,amsfonts,bm,xparse}
\newtheorem{theorem}{Theorem}

\newtheorem{lemma}{Lemma}
\newtheorem{problem}{Problem}
\newtheorem{proposition}{Proposition}
\newtheorem{game}{Game}

\usepackage{stfloats}
\usepackage{cases}
\newtheorem{defn}{Definition}%[section]
\usepackage{algorithmic}
\usepackage{graphicx}
\usepackage{epstopdf}
\usepackage{textcomp}
\usepackage{xcolor}
\usepackage{url}
\usepackage{enumitem}
\usepackage{url}
\usepackage{bbm}
\usepackage{tcolorbox}
\setlist[itemize]{leftmargin=*}
\setlist[enumerate]{leftmargin=*}
\usepackage{multirow}
 
\usepackage{subfigure}
\usepackage[linesnumbered, ruled]{algorithm2e}
\def\BibTeX{{\rm B\kern-.05em{\sc i\kern-.025em b}\kern-.08em
    T\kern-.1667em\lower.7ex\hbox{E}\kern-.125emX}}
\begin{document}
\title{Joint Participation Incentive and Network Pricing Design for Federated Learning
}
\IEEEoverridecommandlockouts
\author{Ningning Ding, 
	Lin Gao, and
	Jianwei~Huang 
	\thanks{This work is supported in part by the National Key Research and Development Program of China (Grant No. 2021YFB2900300), the National Natural Science Foundation of China (Grant 62271434 and 61972113), Shenzhen Science and Technology Program (Projects JCYJ20210324120011032, JCYJ20190806112215116, KQTD20190929172545139, and ZDSYS20210623091808025), Guangdong Basic and Applied Basic Research Foundation (Project 2021B1515120008), Guangdong Provincial Key Laboratory of Big Data Computing, and the Shenzhen Research Institute of Big Data. 
	}				
	\thanks{Ningning Ding is with the Department of Electrical and Computer Engineering, Northwestern University, Evanston, IL 60208, USA and Shenzhen Research Institute of Big Data,  Shenzhen 518172, China.}	
	\thanks{Lin Gao is with the School of Electronics and Information
		Engineering and the Guangdong Provincial Key Laboratory of
		Aerospace Communication and Networking Technology, Harbin Institute of Technology, Shenzhen, China. }	
	\thanks{Jianwei Huang is with School of Science and Engineering, Shenzhen Research Institute of Big Data, The Chinese University of Hong Kong, Shenzhen, Shenzhen 518172, China (corresponding author, e-mail: jianweihuang@cuhk.edu.cn).}		
}
%\and
%\and
%\and
%\and
%\and

\maketitle
\begin{abstract} 
Federated learning protects users' data privacy through sharing users' local model parameters (instead of raw data) with a server. However, when massive users train a large machine learning model through federated learning, the dynamically varying and often heavy communication overhead can put significant pressure on the network operator. The operator may choose to dynamically change the network prices in response, which will eventually affect the payoffs of the server and users. This paper considers the under-explored yet important issue of the joint design of participation incentives (for encouraging users' contribution to federated learning) and network pricing (for managing network resources). Due to heterogeneous users' private information and multi-dimensional decisions, the optimization problems in Stage I of  multi-stage games are non-convex. Nevertheless, we are able to analytically derive the corresponding optimal contract and pricing mechanism through proper transformations of constraints, variables, and functions, under both vertical and horizontal interaction structures of the participants. We show that the vertical structure is better than the horizontal one, as it avoids the interests misalignment between the server and the network operator. Numerical results based on real-world datasets show that our proposed mechanisms decrease server's cost by up to 24.87\% comparing with the state-of-the-art benchmarks.
\end{abstract}
%\vspace{-0.5mm}
%\vspace{-2mm}
\section{Introduction}
%\vspace{-1.5mm}
\subsection{Background and Motivations}
%\vspace{-1mm}
With the fast development of machine-type communications to support the Internet of Things (IoT),  user  devices are generating   unprecedented amount of data\footnote{IoT big data statistics show that the amount of data generated by IoT devices is expected to reach 73.1 ZB (zettabytes) by 2025, corresponding to more than 300\% growth over the 2019 output \cite{enoudata}.} to power intelligent machine learning models \cite{goudos2017survey}. % in many aspects of our life. 
%with increased adoption, IoT devices will globally generate exponentially more data in the following years. The numbers will reach 73.1 ZB by 2025, which equals 422% of the 2019 output
However, users' privacy concerns  often make it risky (or even illegal) to centrally collect and store all users’ data  for model training. 
This motivates the deployment  of federated learning, which  enables effective collaborative  learning while protecting users’ data privacy. 
%A typical cross-device federated learning application platform (e.g., Gboard or automatic driving in Internet of Vehicles) usually consists of (i) a large population of users who use their  local data to collaboratively train a shared learning model, and (ii) a central server who coordinates the training. %Federated learning allows the server to   train a  learning model by using users' private data but without invasion of users' privacy. 
%%Specifically,  in each round of a synchronous training process, each user computes the  parameters of the local learning model based on his local  data and sends his computed model  parameters to the server;  the server  updates the global model based on the users' inputs and  feeds  the  aggregated global model back to  users for their computation in the next round. Users and the server repeat  this process  until achieving  a desired  accuracy, e.g., the training error is smaller than a target threshold  \cite{mcmahan2016communication}. 
%%Different from a  traditional centralized model training process where the central server acquires and stores users' raw data, 
%Federated learning 
During the model training process,   distributed users  keep their private data on their own devices and only share  intermediary model parameters with the central server \cite{mcmahan2016communication}.

Although promising, federated learning still has several under-explored performance bottlenecks, including lack of incentives for participation and heavy communication overhead \cite{kairouz2021advances}. Most existing studies made the optimistic assumption that users are willing to participate in  the federated learning training process (e.g., \cite{Kang2019}). This is not always possible if the users do not receive enough incentives (rewards) to compensate their computation and communication costs \cite{tran2019federated}. 
%That may not be realistic without proper incentives, as users incur various costs during the training process  or may not be interested  in the trained model. 
Although some important earlier works explored the incentive mechanism design for federated learning, they did not consider the large and dynamically changing communication overhead and the network operator's resource pricing (e.g., \cite{sarikaya2019motivating,feng2019joint,zhan2020learning,ding2020optimal,Deng2021}).
%However, the large and dynamically changing communication demand  significantly affects the incentive mechanism design  of the federated learning system \cite{Nguyen2021}.
%While most literature related to the communication of federated learning focused on  improving communication efficiency or reducing communication overhead (e.g., \cite{konevcny2016federated,Saputra2019,Luping2019}), the economic impact of communication overhead (in particular when other services share the same network resources)  is still  under-explored.

Two aspects lead to the heavy and dynamically varying  communication overhead. %, which has significant impacts on the overall network performance.
First, federated learning applications involving a large number of edge devices\footnote{Business Insider Intelligence expects  {vehicles} in the Internet of Vehicles  systems  to rise from 33 million in 2017 to over 77 million by 2025  \cite{insider2}. %Even at lower levels of autonomy, connected vehicles generate around 25 Gigabytes of  {data} per hour \cite{tuxera}.	%The accessibility of sensors and camera modules is making the car industry increasingly data-driven, 	%and self-driving vehicles can come with a myriad of sensors creating machine-to-machine data at the rate of 1GB per second \cite{computerworld}.
}  increasingly involve complex deep neural networks (DNNs) (e.g., Gboard \cite{gboard} and federated automatic driving \cite{nguyen2021deep}). The model parameter update uploaded by each user consists of a large size of gradient vector\footnote{As the state-of-the-art image classification model, Google's NASNet achieves over 80\% accuracy on ImageNet but has a 355MB size \cite{zoph2018learning}. %https://blog.dataiku.com/making-neural-networks-smaller-for-better-deployment-solving-the-size-problem-of-cnns-using-network-pruning-with-keras
}, leading to heavy communication overhead \cite{Luping2019}. %Furthermore,  %users model uploading time is  uncertain and different, due to their unreliable network connections and private time availability
Second, %mobile users often experience dynamically changing network connectivities, which affect the uploading time of the training updates. 
as  mobile users experience different and dynamically changing network connectivities, they may choose to upload their model updates in different time slots (i.e., asynchronously) \cite{mcmahan2016communication,Zhang2020}. This makes the total communication overhead of the federated learning system dynamically changing over time \cite{estiri2021attentive}.  
%Consider the typical example of Internet of Vehicles (IoV), where there is a large number of   users (vehicles)\footnote{Business Insider Intelligence expects  {connected vehicles} in the Internet of Vehicles (IoV) systems  to rise from 33 million in 2017 to over 77 million by 2025  \cite{insider2}. %Even at lower levels of autonomy, connected vehicles generate around 25 Gigabytes of  {data} per hour \cite{tuxera}.
%	%The accessibility of sensors and camera modules is making the car industry increasingly data-driven, 
%	%and self-driving vehicles can come with a myriad of sensors creating machine-to-machine data at the rate of 1GB per second \cite{computerworld}.
%}. 
%If they jointly train a federated learning model, there will be a huge amount of information exchange and model transmission, especially for training the big deep learning models. 
As a result, such heavy and dynamic communication demand can significantly influence the  network operator’s  resource usage and pricing strategy, which  in turn affect users' incentives to join the federated learning system \cite{Nguyen2021}. %experiences and network  payment as well as  server's incentive designed for users 
 %, which Without the consideration of communication cost, not only affect the invetive (user cost), but also the NO's profit and network cost

To overcome the above bottlenecks, we focus on the  joint design of incentive mechanism  and network pricing in federated learning, with several challenges to tackle. 
First, the dynamic  resource demand motivates the network operator to set dynamic prices   to manage the network quality and  maximize its profit,  which in turn will change the resource demand distribution over time (e.g., a high price    discourages users'  usage)\cite{Wang2019}.
Moreover, each user's network usage  affects other users' payoffs through network congestion (i.e., network externality).
%It is challenging for service providers to take all customers’ demands and strategies into consideration for designing the optimal prices.
 It is difficult for the network operator to design the optimal prices considering all heterogeneous users' different usage choices and their network externality.
Second,  users' private information (e.g., training costs)  increases the difficulty for server's incentive design and network operator's  pricing.  Selfish users  can misreport their information for a more favorable outcome  \cite{ding2020optimal}. 
Third,   the complex interaction among users, server, and network operator also significantly affects how they can derive their optimal  strategies. Specifically, there are two widely-considered interaction structures in the market depending on participants' relative market powers \cite{ghosh2018pricing,infocom22,syverson2019macroeconomics}:
\begin{itemize}
	\item \emph{Horizontal interaction structure:} the network operator and federated learning server  announce their pricing and incentive mechanisms simultaneously, based on which users   make  participation decisions  (to be introduced in Section \ref{twog}). 
	\item \emph{Vertical interaction structure:} the network operator, server, and users  make their decisions sequentially  (to be introduced in Section \ref{threeg})\footnote{For example, Cosmo (selling smartwatches) and Things Mobile (selling smartwatch SIM cards) have similar market power, and they usually forms a horizontal  interaction structure; China Mobile has  larger market power than BYD Auto, and they usually forms a vertical structure \cite{infocom22}.}.
\end{itemize}
\vspace{-0.5mm}
Different structures require   different incentive and pricing considerations. When it is feasible for both structures to exist in a market, it is also important to compare the performance of these structures and provide policy guidelines regarding which one is more beneficial to the society.  

These challenges  motivate us to answer the following interesting questions in a federated learning system:

\noindent\textbf{Key Question 1:}  \emph{What is the server's optimal incentive mechanism for heterogeneous users with  private information, considering the heavy communication overhead?}

\noindent\textbf{Key Question 2:}  \emph{How should the network operator set the prices to maximize its profit, considering the dynamically changing network resource demand?}
%2) What is the network operator’s optimal dynamic pricing for its network service at different times, considering the communication demand caused by federated learning?
	%\item What are heterogeneous users'   optimal training participation strategies and network usage strategies?% users' mutual influence?
	
\noindent\textbf{Key Question 3:}  \emph{Which interaction structure is better in terms of  the payoffs of  the server, the network operator, and users?}

\vspace{-0mm}
\subsection{Contributions}
\vspace{-1mm}
We summarize our key novelty and   contributions below.
\begin{itemize}
	\item  \emph{Incentive mechanism design considering dynamically changing  network resource demand}. To the best of our knowledge, this is	the first analytical study on incentive mechanism design 	for federated learning considering dynamic network resource demand. Such a study is practically important for the sustainable development of federated learning systems.
	
	\item \emph{Joint design of optimal contract and network  pricing under different interaction structures}.
	We propose multi-stage games to analyze the  server's optimal contract and  the network operator's optimal dynamic  pricing, under both horizontal and vertical interaction structures. With heterogeneous users'  private information, the optimization problems  of the network operator and the server are non-convex  and of a high complexity (e.g., a large number of constraints). We solve these problems by converting the constraints into simpler but equivalent ones and properly  transforming  variables and functions to obtain convex or analyzable problems.%, and decomposing the multi-variable optimization into sequential single-variable optimizations. 
	\item  \emph{Comparison of interaction structures}. We show that the vertical interaction structure is better than the  horizontal  structure for users,   server, and   network operator. This is because the sequential decision process  under the vertical structure  avoids the scenarios where users incentivized by the server   cannot afford the network payment.
	\item \emph{Insights about network pricing and demand distribution}. When users are congestion-tolerant, we show that   it is optimal for the network operator to achieve a water-filling network demand distribution and set the same price for the time slots chosen by at least one user. However, when users are congestion-sensitive, time slots with less background network demands encourage the federated learning users' selection  but still have less total network demands. This is because the network operator needs to consider users' total congestion cost in each selected time slot.	
%	the optimal prices and total network usage  are different in different chosen time slots and depend on users' congestion costs. 
	\item \emph{Performance evaluation}. Numerical results based on   real-world datasets show that   our proposed  mechanisms decrease the server's cost by up to 24.87\% and	increase the network operator's profit  by up to 1245.25\%, compared with the state-of-the-art benchmarks.	
\end{itemize}

\vspace{-1mm}
\subsection{Related Work}
\vspace{-1mm}
\label{literature}
%\subsubsection{Federated Learning}
Most studies on federated learning focused on   improving training  efficiency  (e.g., \cite{ren2020accelerating,luo2021cost}),  enhancing security (e.g.,  \cite{fung2018mitigating,tan2020toward}), and preserving privacy (e.g., \cite{hao2019towards,sun2022profit}). Most of the results were derived  under an optimistic assumption that users are willing to participate in   federated learning, which may not be realistic without proper incentives.% to the users.

A carefully designed incentive mechanism  can  elicit users' truthful information, promote cooperation, and enhance system efficiency in federated learning \cite{kairouz2021advances}. Although  federated learning has seen increasingly more applications in practice, there are only a few important earlier works on the incentive mechanism design (e.g., \cite{sarikaya2019motivating,feng2019joint,zhan2020learning,ding2020optimal,Deng2021,jiao2020toward,Kang2019}), with a few limitations. For example, Sarikaya et al. in \cite{sarikaya2019motivating} studied a complete information scenario where the server knows the private information of users.  % some literature did not consider the information asymmetry among the participants (e.g., \cite{sarikaya2019motivating}). %For example,  Sarikaya et al.  assumed a complete information scenario where the server knows the private information of users (e.g., costs). 
%Third, few work considered the non-IID data scenario (e.g., \cite{jiao2020toward}).
 % (e.g., \cite{feng2019joint,zhan2020learning}). %, and the corresponding solutions are not easily generalizable  when users have private information 
Kang et al. in \cite{Kang2019} and Jiao et al. in \cite{jiao2020toward} focused on the  incentive mechanism design under incomplete information yet without  closed-form solutions. % (e.g., \cite{jiao2020toward,Kang2019}). 
Feng et al. in \cite{feng2019joint} and Zhan et al. in \cite{zhan2020learning} modeled users' independent communication costs, without considering users' mutual influence of network usage (e.g., network congestion). 
Building upon these earlier work, we propose a  more general and practical model with private information and users'  network externality. %,  and provide a more comprehensive mechanism design. % with closed-form solutions, and consider the effect of information asymmetry.

More importantly, our work has  two key novelties compared with  prior studies on incentive design for federated learning.  First, to the best of our knowledge, prior related literature   did not consider the impact of  dynamic  network resource demand, which is challenging to analyze yet practically significant.  Instead of only focusing on the interaction between the server and users, our work will perform the joint optimization of network operator's resource pricing and server's incentive design for users.
Second, no prior work studied incentive mechanism for federated learning under different interaction structures. However, this is important for both the system participants and policy makers.  %Participants in different   structures have different interactions and different information of others' strategies, leading to different incentive mechanisms.  
%Thus, the two aspects substantially increase the challenge of  incentive mechanism design.
 %Table \ref{tablit}  summarizes the key differences between our work and the related literature. 

%
%
%\subsubsection{Network Resource Pricing}
%vertical structure has been widely used in data pricing and horizontal structure is new and under explored. 
%
%...

The rest of the paper is organized as follows. We first introduce the system model   in Section \ref{model}. We then study the optimal incentive mechanism and network pricing design under the  vertical  structure and the  horizontal structure  in Sections \ref{v2} and \ref{h2}, respectively.  We present some interesting insights about  congestion-tolerant  users in Section \ref{h1}.  We perform simulations based on real-world datasets in Section \ref{sim} and conclude in Section \ref{conclusion}.

\vspace{-1mm}
\section{System Model}
\vspace{-0.5mm}
\label{model}
We consider a typical federated learning platform (e.g., federated automatic driving   in the Internet of Vehicles (IoV)), where the model training is distributed over $I$ users and coordinated by a central server. The communication (i.e., model updates transmission) between  users and the server  during the model training is supported by a mobile network operator (e.g., AT\&T). In the following, we will first introduce the federated learning process, then specify the strategies and payoffs of users, server, and network operator, and finally formulate the games among these participants.
\vspace{-2mm}
\subsection{Federated Learning Process}
\vspace{-1.3mm}
\label{fl}
Federated learning is a distributed machine learning paradigm, in which many users   collaboratively train a shared learning model under a server's coordination.
%{\color{blue}example of IoV?} 
%As an illustrative  example, Gboard is a Google keyboard software which relies federated learning to help  users predict the next word (to be typed) based on the current word (that has been just typed). %Since the typing data from each mobile user is limited, Gboard relies the data from millions of users to achieve an effective prediction.  It asks  users to use their  local data about input behaviors to cooperatively train a global learning model. Each user only needs to share model parameters with the server without uploading his raw data. 
Consider an example of data $(x_a,y_a)$, where $x_a$ is the input (e.g., an image) and $y_a$ is the label (e.g., the object  in the image). The objective of learning is to find the proper model parameter $w$ that can predict the label $y_a$ based on the input $x_a$. Let us denote  the prediction value as $\tilde{y}(x_a;w)$. The gap between the prediction  $\tilde{y}(x_a;w)$ and the ground truth label $y_a$ is characterized by the  prediction loss function $f_a(w)$. %For example, the loss function in linear regression is $f_{i}(w)=\frac{1}{2}\left(x_{i}^{T} w-y_{i}\right)^{2}, y_{i} \in \mathbb{R}$. %; the loss function in support vector machine is $f_{i}(w)=\left\{0,1-y_{i} x_{i}^{T} w\right\}, y_{i} \in\{-1,1\}$. 
If user $i$ uses a set $\mathcal{S}_i$ of local data with data size $s_i$ to train the model,  the loss function of user $i$  is the average prediction loss on all his data $a \in \mathcal{S}_{i}$: 
\vspace{-3mm}
\begin{equation}
F_{i}(w)=\frac{1}{s_{i}} \sum_{a \in \mathcal{S}_{i}} f_{a}(w).
\end{equation}
\vspace{-4.5mm}

\noindent The purpose of federated learning is to compute the model parameter $w$ by using all users' local data. The optimal model parameter $w^*$  minimizes  global loss function, which is a weighted average of all users' loss functions:
\vspace{-3.5mm}
\begin{equation}
\label{weight}
w^*=\arg\min _{w} f(w)=\arg\min _{w}\sum_{i=1}^{I} \frac{s_{i}}{s} F_{i}(w),
\end{equation}
\vspace{-4mm}

\hspace{-3.5mm}where $s$ is the total data size of all users \cite{mcmahan2016communication}.  
%Note that this decentralized approach is equivalent to directly minimizing the average loss function of the whole data, i.e., 
%$
%\min _{w \in \mathbb{R}^{d}} f(w)  \stackrel{\text { def }}{=} \frac{1}{n} \sum_{i=1}^{n} f_{i}(w)
%$, 
%where $n$ is the total number of data examples of users.

%Specifically,  in each round of a synchronous training process, each user computes the  parameters of the local learning model based on his local  data and sends his computed model  parameters to the server;  the server  updates the global model based on the users' inputs and  feeds  the  aggregated global model back to  users for their computation in the next round.
We consider  the widely adopted synchronous federated learning  that proceeds in rounds of communication. Each global training round starts when the server broadcasts the current global model parameter to all users and ends after all users upload their local parameter updates to the server for aggregation  (so that the server can produce a new global model parameter). % (as shown in Algorithm \ref{alg:A} \cite{mcmahan2016communication}).  
The key advantage of the synchronous algorithms is that they have provable convergence  (e.g., \cite{lim2019federated,kairouz2019advances}). %A typical synchronous federated learning algorithm works as . 

Next, we   model the  strategies of the network operator, server, and  users in each training  round.

\vspace{-1.7mm}
\subsection{Time Frame, User Type, and Strategies}
\vspace{-1mm}
\subsubsection{{Time Frame}}
We refer to one training round as  one time frame, which is divided into $\mathcal{T}\triangleq\{1,2,...,T\}$ time slots. 
For example,  the time frame can be one day,  which can be further divided into 24 time slots (each with one hour).
% As long as the users upload results within the deadline, it make no difference whether they upload earlier or later
In our model, we focus on the optimization in one typical time frame.
\subsubsection{User Type}
We consider a set $\mathcal{I}\triangleq\{1,2,...,I\}$ of users in the federated learning system. Users are distinguished by their marginal data-usage costs  $\theta$. We refer to a user with $\theta_j$ as a type $j$ user. Without loss of generality,  $I$ users belong to a set $\mathcal{J}\triangleq\{1,2,...,J\}$ of $J$ types. Each type $j$ has $I_j$ users, with $\sum_{j \in \mathcal{J}} I_{j}=I$. We assume that $\theta_1< \theta_2 <...<\theta_J$, and the maximum data size that a user can generate is $d^{\max}$.  
The total number of users $I$ and the specific number of each type $I_j$ are public information, but each user's type is private information\footnote{It is easy for the server and the network operator to have the knowledge about statistics of type information through market research and past experiences, but it is hard to know each user's private type \cite{wang2018multi}.}.

\subsubsection{{Network Operator's Pricing}}
The  network operator has the flexibility of setting  different  network prices $\boldsymbol{p}\triangleq 	\{p(t)\}_{t\in \mathcal{T}}$ in different time slots $t\in \mathcal{T}$. Due to regulatory concerns, the maximum price for any time slot will be $p_0$. 
%We denote the maximum   price that the network operator can set as $p_0$, which is determined by regulations. % (i.e., $p(t)\le p_0, \forall t\in \mathcal{T}$). %The network operator can only provide discounts, but cannot charge prices higher than the benchmark (i.e., $p(t)\le p_0, \forall t\in \mathcal{T}$)\footnote{The constraint of providing discounts ensures that the new pricing scheme can only reduce the cost of the users, and hence will be embraced by users and supported by regulators during actual implementation. Moreover, since not providing any discounts is a feasible choice, the network operator will not experience a total cost higher than today’s time independent pricing benchmark. Hence the “discount-only” pricing scheme leads to a win-win situation for both the network operator and users.}. 

%Note that it is practical that the network operator sets the same price in the same time slot of everyday as the daily network demand distribution is usually similar \cite{bergroth202224}. Also, the server collects vehicles' local model once a day after users finish their daily driving. Therefore, we only need to optimize all participants'  payoffs/profits in each time frame (one day here). The corresponding strategies   are also the long-term optimal strategies which maximize their profits or payoffs. 

\subsubsection{{Server's Contract}}
The server wants  users to upload their local training results by the end of the time frame and contribute  as much data for local training as possible. 
The server will design a contract\footnote{Both contract theory  and auction theory are promising and widely adopted theoretic tools for dealing with incentive  problems with private information. Contract  is more applicable to the case where the server knows user type distribution  but does not know each user's type, while auction  is more suitable when the server  does not even know the user type distribution \cite{bolton2004contract,krishna2009auction}. } $\boldsymbol{\phi}\triangleq \left\{\phi_{j}\right\}_{j \in \mathcal{J}}$, which contains $J$ contract items (one for each user type). Each contract item $\phi_{j} \triangleq\left(d_{j}, r_{j}\right)$ specifies the relationship between each type-$j$ user’s data size (for local training)  and reward. Here $d_j$ is the required training data size  in each training round, and $r_j$ is the corresponding reward if a  type-$j$ user  completes his training task by the end of the current data frame (i.e., within the current training round)  with required   data size.

\subsubsection{Users' Choices}
Each user    decides whether to participate in the training, (if yes) which contract item to choose, and which time slot  to upload the training results. %\footnote{We consider that  one time slot is enough for a user to train and upload his local model (no matter how much data he uses).  %In the future work, we can extend to the scenario where more data  leads to longer training time, which may exceed one or more time slots.
%}. 
The choice of different time slots may lead to different network congestion costs and network prices for users.
A user  will not participate if his payoff (defined in Section \ref{up}) is negative.

\vspace{-1mm}
\subsection{Payoffs and Profits}
\vspace{-0.5mm}
\subsubsection{{Users}}
\label{up}
Since a user's type is private information, he can choose a contract item not designed for his type. 
When a  user $i \in \mathcal{I}$   chooses the contract item $\phi_{j}$ and the time slot $t_i$, his payoff  is the difference between the reward from the server and the costs (on his data usage, network payment, and network congestion):
\vspace{-2.5mm}
\begin{equation}
W_U^i(\phi_{j},t_i)=r_j\hspace{-0.3mm}-\theta_i d_j-p(t_i)\hspace{-0.3mm}-\hspace{-0.3mm}\beta\hspace{-0.3mm} \left(\sum_{k\in \mathcal{I}}\mathds{1}_{t_k=t_i}+h(t_i)\right)^2\hspace{-1.5mm},
\end{equation}

\vspace{-2.5mm}
\noindent where $\sum_{k\in \mathcal{I}}\mathds{1}_{t_k=t_i}$ is the normalized network usage in this system (i.e., the number of federated learning users who choose to upload model parameters in  the same  time slot $t_i$ as user $i$), $h(t_i)$ is the  network usage from other systems  at time slot $t_i$ (i.e., background network usage at time slot $t_i$), and 
$\beta \left(\sum_{k\in \mathcal{I}}\mathds{1}_{t_k=t_i}+h(t_i)\right)^2$ is the congestion cost. The quadratic congestion cost  captures the increasing marginal cost feature of congestion-sensitive users\footnote{For example, due to the  high requirements on  network quality, users in an IoV system can be very sensitive to the network congestion, especially when the congestion is serious. We will study the congestion-tolerant users in Section \ref{h1}.} \cite{Burra2019,Kalashnikov2018}.
%Note that we consider all users network usage for uploading results are the same, as they train the same federated learning model. We normalized each user's network usage to be one and scale the value of $h(t)$  accordingly.

\subsubsection{{Server}}
The server's cost is determined by the accuracy loss of the global model and the total rewards for users\footnote{We consider that the server's network cost for broadcasting the model at the beginning of the training round is a constant.  Mathematically, it will not affect the optimization and analysis in this paper, so we do not model it here.}.

First, we characterize the expected  accuracy loss of the global model.   The model accuracy loss after $D$ training rounds is measured by  the difference between the   prediction loss with parameter $w_D$  and that with the optimal parameter $w^{*}$, i.e., $f(w_D)-f(w^*)$ (defined in Section \ref{fl}). The expected difference  is bounded by  $O(1 / \sqrt{BD}+1/D)$ \cite{li2014efficient,dekel2012optimal}, where $B$ is all users' total training data size in each round, i.e., $B=\sum_{j\in \mathcal{J}}I_jd_j$. Given a fixed $D$ (as we optimize the each-round accuracy), minimizing the accuracy loss bound is  mathematically equivalent to minimizing $1/{\sqrt{\sum_{j\in \mathcal{J}}I_jd_j}}$.\footnote{Note that optimizing the single-round accuracy will also guarantee the performance of the entire training process, as it is equivalent to minimizing the expected accuracy loss given any total training round $D$, i.e., $1/{\sqrt{D\sum_{j\in \mathcal{J}}I_jd_j}}+1/D$.  Mathematically, the constant $D$ here will not affect the optimization. Moreover, we will use experimental accuracy loss in the simulations in Section \ref{sim} to validate our analytical results.} It is clear that the model accuracy loss  decreases in users' total training data size.
% In our setting, we optimize the each-round accuracy, i.e., $D$ is fixed and $B$ is all users' total training data size. Thus, the server's expected accuracy loss is $1/{\sqrt{D\sum_{j\in \mathcal{J}}I_jd_j}}+1/D$, which can be simplified as $1/{\sqrt{\sum_{j\in \mathcal{J}}I_jd_j}}$ as mathematically the constant $D$ will not affect the following optimization.
%{\color{blue}(remove the above and just say according to some literature, accuracy loss has a square root relationship with the total data size)}

Then, we consider the server's total rewards for all  users. If all users choose to participate in the contract and choose their corresponding  contract items\footnote{As we shall see in Section \ref{v2}, without loss of generality,  we will design the contract to ensure that each user will choose the contract item designed for his type  (i.e., incentive compatibility).}, the total rewards is $\sum_{j\in \mathcal{J}}I_jr_j$. 

To summarize, the server's cost  is:
\vspace{-1mm}
\begin{equation}
W_S=\frac{1}{\sqrt{\sum_{j\in \mathcal{J}}I_jd_j}} +\xi \sum_{j\in \mathcal{J}}I_jr_j,
\end{equation}

\vspace{-1.5mm}
\noindent where  $\xi$ is server's weight on the rewards.  A larger $\xi$ means that the server is more concerned about minimizing the reward and less concerned about minimizing the accuracy loss.

\subsubsection{{Network Operator}}
Network operator's  profit  is the difference between the revenue from users and the total cost for providing network service in all time slots of a   time frame:
\vspace{-1.5mm}
\begin{equation}
W_O=\sum_{i\in \mathcal{I}}p(t_i)-\gamma \sum_{t\in \mathcal{T}} \left(\sum_{i\in \mathcal{I}}\mathds{1}_{t_i=t}+h(t)\right)^2.
\end{equation}

\vspace{-1.5mm}
\noindent Here the quadratic network cost captures the widely considered increasing marginal cost feature (e.g., \cite{tahiri2018reservoir,Doan2017}). Intuitively, network operator's cost monotonically increases in the network usage amount (which includes the network usage in this system and the background usage from other systems). However, due to the network operator' limited network resource,  when there is already a huge amount of network usage, further increasing the usage will lead to  even more significant costs. 
The $\gamma$ indicates network operator's weight on the network cost. A larger $\gamma$ means that the server is more concerned about minimizing the network cost and less concerned about maximizing the revenue. 

%The optimization of their profits and payoffs also depends on the interactions among users, server, and network operator.

\vspace{-1mm}
\subsection{Game and Interaction Structure}
\vspace{-0.5mm}
\label{game}
We focus on two widely-considered practical interaction structures:  vertical and horizontal  structures \cite{ghosh2018pricing,infocom22}. Different structures correspond to different game formulations.
\subsubsection{Three-Stage Game (Vertical Structure)}
\label{threeg}
\begin{figure}
	\centering
	\includegraphics[width=3.2 in]{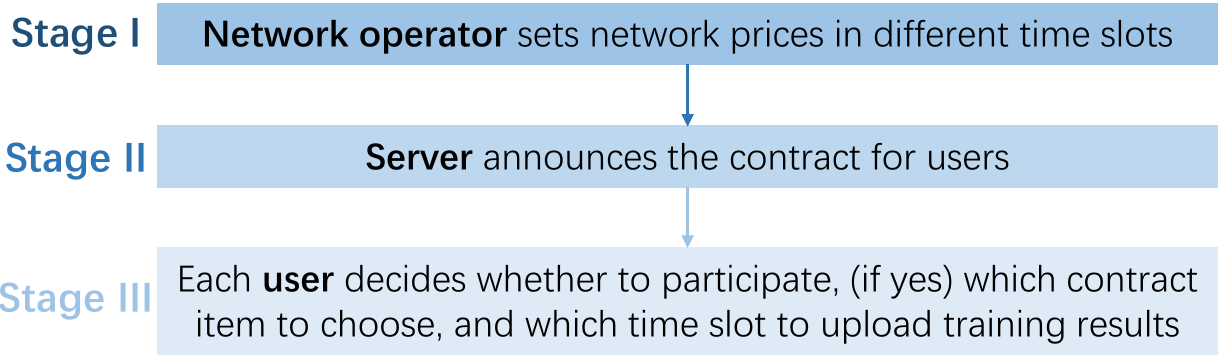}
	\vspace{-3mm}
	\caption{Vertical interaction structure.}
		\vspace{-3.7mm}
	\label{three}
\end{figure}
As shown in Fig.~\ref{three}, we model the interaction among the participants as a three-stage Stackelberg game under the vertical structure. The network operator sets the network prices in different time slots in Stage I.  After observing the network prices, the server  announces the contract for users in Stage II. Given the network operator's prices and the server's contract, each user decides whether to participate and (if  yes) chooses the contract item  and the time slot  for uploading training results in Stage III. % Moreover,  there is also  a non-cooperative game among users  in Stage III, as each user aims to maximize his own payoff. 
%We will use the backward induction to analyze the game in Section \ref{v2}. %Specially, in Stage I, we will first calculate the best responses of network operator and the server, and then obtain their optimal strategies at the equilibrium by combining their best responses.
\subsubsection{Two-Stage Game (Horizontal Structure)}
\label{twog}
\begin{figure}
	\centering
	\includegraphics[width=3.2 in]{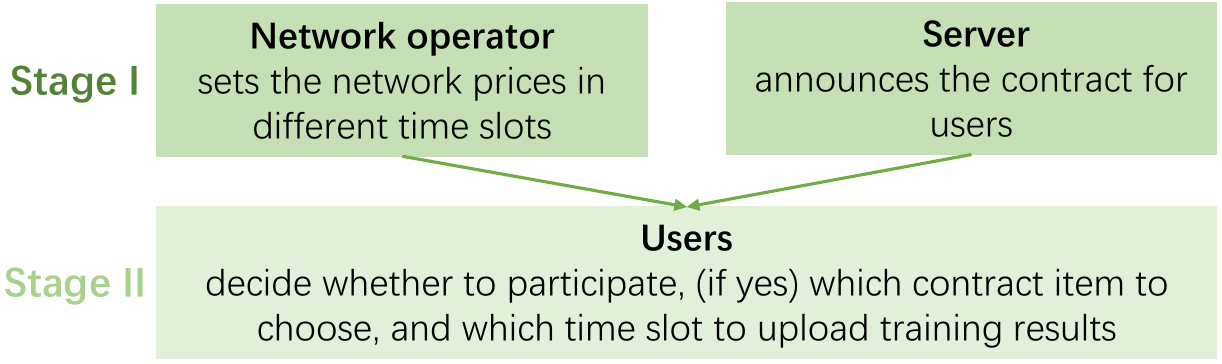}
	\vspace{-3mm}
	\caption{Horizontal interaction structure.}
	\vspace{-6.5mm}
	\label{two}
\end{figure}
As shown in Fig.~\ref{two}, we model their interaction   as a two-stage Stackelberg game. In Stage I, the network operator sets the prices in different time slots, and meanwhile the server announces the contract for users. %Thus, there is  a non-cooperative game between the network operator and the server in Stage I. 
In  Stage II, users then make their decisions, which also leads to a non-cooperative game.
%Moreover, in Stage I, there is a non-cooperative game between the network operator and the server; , there is also a non-cooperative game among users. 
%Specially, in Stage I, we will first calculate the best responses of network operator and the server, and then obtain their optimal strategies at the equilibrium by combining their best responses.

Next,   we will use backward induction to analyze these two interaction structures in the next two sections.  %the three-stage game for the vertical structure in Section \ref{v2} and the two-stage game for the horizontal structure in Section \ref{h2}.

\section{Incentive Mechanism and Network Pricing Design in Three-Stage Game  (Vertical Structure)}
\label{v2}
In this section, we study the optimal incentive mechanism design and network resource pricing under the vertical interaction structure. Specifically, we  first study users' optimal strategies  in Section \ref{users},   then   calculate the  server's optimal contract  in Section \ref{servers}, and finally   derive the network operator's optimal pricing in Section \ref{networko}.
\vspace{-1mm}
\subsection{Users' Optimal Strategies in Stage III}
\vspace{-0.5mm}
\label{users}
We first formally define users' non-cooperative   game in Stage III and the corresponding equilibrium as follows.

\vspace{-1mm}
\begin{tcolorbox}
	\begin{game}[Stage III: Users' Game under the Vertical Structure]%\quad \\
		\label{game1}
		%	\vspace{-4.5mm}
		The game among users in Stage III is
		\begin{itemize}
			\item Players: $I$ users in set $\mathcal{I}$.
			\item Strategy space: each user $i\in \mathcal{I}$ decides whether to participate, which  contract item $\phi_{i}\in \boldsymbol{\phi}$ to choose, and which time slot $t_i\in \mathcal{T}$ to upload his model parameters. 
			\item Payoff function: each user $i\hspace{-0.5mm}\in\hspace{-0.5mm} \mathcal{I}$ maximizes his payoff 
		\end{itemize}
	\begin{equation}
	\begin{split}
\hspace{-2.5mm}W_U^i&	(\phi_{i},t_i;\boldsymbol{\phi_{-i}},\boldsymbol{t_{-i}})=\\&r_i\hspace{-0.5mm}-\hspace{-0.5mm}\theta_i d_i\hspace{-0.5mm}-\hspace{-0.5mm}p(t_i)\hspace{-0.5mm}-\hspace{-0.5mm}\beta\hspace{-0.3mm} \bigg(\hspace{-0.5mm}\sum_{k\in \mathcal{I}}\mathds{1}_{t_k=t_i}\hspace{-0.5mm}+\hspace{-0.5mm}h(t_i)\hspace{-0.5mm}\bigg)^2\hspace{-1mm},
	\end{split}
	\end{equation}
	\vspace{-3mm}
	
\noindent	where $\boldsymbol{\phi_{-i}}\triangleq\{\phi_{i'}\}_{i'\in \mathcal{I}\backslash\{i\}}$ and $\boldsymbol{t_{-i}}\triangleq\{t_{i'}\}_{i'\in \mathcal{I}\backslash\{i\}}$.
	\end{game}
%\end{tcolorbox}
%	\vspace{-3mm}
%	\begin{tcolorbox}
	\begin{defn}[Equilibrium of Game \ref{game1}]
		The equilibrium of  Game \ref{game1} is a choice profile $\{(\phi_i^*,t_i^*)\}_{i\in \mathcal{I}}$,  such that each user achieves his maximum payoff assuming other users are following the equilibrium strategies, i.e., $\forall \phi_{i} \hspace{-1mm}\in \boldsymbol{\phi},$ $ \forall t_i\hspace{-1mm}\in \mathcal{T},$
		\vspace{-1mm}
		\begin{equation}
		W_U^i(\phi_{i}^*,t_i^*;\boldsymbol{\phi_{-i}}^*,\boldsymbol{t_{-i}}^*)\ge  W_U^i(\phi_{i},t_i;\boldsymbol{\phi_{-i}}^*,\boldsymbol{t_{-i}}^*).  
		\end{equation}
	\end{defn} 
\end{tcolorbox}
\vspace{-1mm}
%The optimal $t_i^*$ for each user $i$ satisfies
%\begin{equation}
%U_i-p(t_i^*) -\beta \left(\sum_{k\in \mathcal{I}}\mathds{1}_{t_k=t_i^*}+h(t_i^*)\right)^2 \ge U_i-p(t) -\beta \left(\sum_{k\in \mathcal{I}}\mathds{1}_{t_k=t}+h(t)\right)^2, \;\forall t\in \mathcal{T},
%\end{equation}
%which is equivalent  to
%\begin{equation}
%\label{7}
%p(t_i^*)+\beta \left(\sum_{k\in \mathcal{I}}\mathds{1}_{t_k=t_i^*}+h(t_i^*)\right)^2 \le p(t) +\beta \left(\sum_{k\in \mathcal{I}}\mathds{1}_{t_k=t}+h(t)\right)^2, \;\forall t\in \mathcal{T}.
%\end{equation}
By solving  Game \ref{game1}, we have the following lemma:
\begin{lemma}
	\label{th4}
	%The equilibrium of Game \ref{game1} exists if and only if 
	At the equilibrium of Game \ref{game1}, all chosen time slots $\{t_i^*\}_{i\in \mathcal{I}}$   have identical lowest user network cost: %each participating user $i$ will choose a time slot  with the lowest network cost: 
%	It is optimal for each user $i$ to choose a time slot  with the lowest network cost:
	\vspace{-1mm}
	\begin{equation}
	\label{choice}
	c(\boldsymbol{p})\triangleq{\min}_{t\in \mathcal{T}} \bigg(p(t)+\beta \Big(\sum_{k\in \mathcal{I}}\mathds{1}_{t_k^*=t}+h(t)\Big)^2\bigg),
	\end{equation}
	and all unselected time slots have network costs larger than $c(\boldsymbol{p})$.
%	\begin{equation}
%	\label{choice}
%	\hspace{-0.2mm}t_i^*\hspace{-0.5mm}(\boldsymbol{\phi},\hspace{-0.3mm}\boldsymbol{p})\hspace{-0.6mm}=\hspace{-0.6mm}{\arg\min}_{t\in \mathcal{T}}\hspace{-0.5mm} \bigg(\hspace{-1mm}p(t) \hspace{-0.3mm}+\hspace{-0.3mm}\beta \hspace{-0.3mm}\Big(\hspace{-0.5mm}\sum_{k\in \mathcal{I}}\hspace{-0.5mm}\mathds{1}_{t_k=t}\hspace{-0.3mm}+\hspace{-0.3mm}h(t)\hspace{-0.5mm}\Big)^2\hspace{-0.5mm}\bigg)\hspace{-0.7mm}, \hspace{-0.5mm} \forall i\hspace{-0.5mm}\in\hspace{-0.5mm} \mathcal{I},
%	\end{equation}
	Each user $i$ will choose the contract item $\phi_{i}^*$ that maximizes his payoff and gives him a non-negative payoff.
%	\vspace{-1mm}
%	\begin{equation}
%	 W_U^i(\phi_{i}^*,t_i^*)\ge  W_U^i(\phi_{k},t_i^*),\forall i,k\in \mathcal{I}.\label{8b}
%	\end{equation} 
%	Users will participate if and only if they can obtain  non-negative payoffs, i.e., 
%	\vspace{-1mm}
%	\begin{equation}
%	W_U^i(\phi_{i}^*,t_i^*)\ge 0,\forall i\in \mathcal{I}.\label{8a}\\
%	\end{equation}
\end{lemma}
\vspace{-0.5mm}
As shown in Lemma \ref{th4}, users will   choose the time slots with the  lowest network cost (i.e., sum of network price and congestion cost) at the equilibrium. Each user can have multiple optimal choices in time slots, each corresponding to a different equilibrium but the same lowest network cost. We denote the lowest network cost as $c(\boldsymbol{p})$, which  depends on the network pricing in Stage I. Moreover, users' contract item choices depend on the server's contract design in Stage II.

%Moreover, choosing a time slot with smaller  price will increase users'  payoffs. However, if many users choose the same time slot, the network operator's resource usage cost at this time slot will be very large, which may increase the network  price for this slot in return.  Therefore, users' network usage at the equilibrium  is related to network operator's dynamic pricing, which we will analyze in Section \ref{brno}. 

%The inequalities   \eqref{8b} and \eqref{8a} indicate that the server needs to design the contract under the \emph{Individual Rationality} (IR) and \emph{Incentive Compatibility} (IC) constraints. Specifically, individual rationality   \eqref{8a}  means that a user will participate if and only if he can obtain a positive payoff, and incentive compatibility   \eqref{8b} means that a user  maximizes his payoff by choosing the contract item intended for him. Next, we will analyze the server's optimal contract design.

\vspace{-1mm}
\subsection{Server's Optimal   Contract    in Stage II}
\label{servers}
\vspace{-0.5mm}
Given the network operator's prices in Stage I, the server needs to design the contract $\boldsymbol{\phi}$ in Stage II,  considering the optimal strategies of users in Stage III.
In this case, the server's optimization problem under the vertical structure is as follows.

\vspace{-1.5mm}
\begin{tcolorbox}
\begin{problem}[Server's Contract Design in Stage II]
	\label{p1}
	\vspace{-1.5mm}
	\begin{equation*}
	\begin{split}
	\min \;&  \frac{1}{\sqrt{\sum_{j\in \mathcal{J}}I_jd_j}} +\xi \sum_{j\in \mathcal{J}}I_jr_j\\
	\rm{s.t.}\; &W_U^i(\phi_{i}(\boldsymbol{\phi}),t_i(\boldsymbol{\phi}))\ge 0,\forall i\in \mathcal{I} \textrm{ (IR)}\\
	&W_U^i(\phi_{i}(\boldsymbol{\phi})\hspace{-0.3mm},t_i(\boldsymbol{\phi}))\hspace{-0.5mm}\ge \hspace{-0.5mm} W_U^i(\phi_{k}(\boldsymbol{\phi})\hspace{-0.3mm},t_i(\boldsymbol{\phi}))\hspace{-0.3mm},\hspace{-0.3mm}\forall i,k\hspace{-0.5mm}\in\hspace{-0.5mm} \mathcal{I}\hspace{-0.5mm}
	\textrm{ (IC)}\\
	&0\le d_j\le d^{\max}, \forall j\in \mathcal{J}\\
	\rm{var.}\; & \boldsymbol{\phi}=\{\left(d_{j}, r_{j}\right)\}_{j\in \mathcal{J}}
	\end{split}
	\end{equation*}
\end{problem}
\end{tcolorbox}
\vspace{-1.5mm}
The   server needs to design the contract under the \emph{Individual Rationality} (IR) and \emph{Incentive Compatibility} (IC) constraints. Specifically, individual rationality     means that a user will participate if and only if he can obtain a positive payoff, and incentive compatibility   means that a user  maximizes his payoff by choosing the contract item intended for him. 

Solving Problem \ref{p1} involves  two challenges. First, users' contract item choices $\phi_{i}$ (related to their marginal training costs $\theta_i$) and upload time choices $t_i$ will both affect users' payoffs and thus the server's optimal strategies, leading to a challenging multi-dimensional contract design. However,   based on Lemma \ref{th4}, we can simplify the analysis into a one-dimensional contract design only about $\theta_i$, as  all participating users' different time slot choices lead to the same network cost $c(\boldsymbol{p})$ (i.e., same impact on their payoffs).
%\vspace{-1mm}
%\begin{equation}
%c(\boldsymbol{p})\triangleq {\min}_{t\in \mathcal{T}} p(t) +\beta \bigg(\sum_{k\in \mathcal{I}}\mathds{1}_{t_k=t}+h(t)\bigg)^2, \forall i\in \mathcal{I}.
%\end{equation}
%
%\vspace{-2mm}
%% Thus, a type-$j$ user's payoff when choosing the contract item $\phi_{j}$ is
%%\begin{equation}
%%W_U(\theta_j,\phi_{j})=r_j-\theta_j d_j-c(\boldsymbol{p})
%%\end{equation}
%%Therefore, the optimal contract is  similar to Theorem \ref{thm3}. We just need to replace $p$ in  Theorem \ref{thm3} with $c(\boldsymbol{p})$ here (as shown in Theorem \ref{thm4}).
%\noindent 
Second, as the total number of IR and IC constraints  is large (i.e., $I^2$), it is challenging to  obtain the optimal contract directly. To overcome such a complexity issue,  we   first transform the constraints into a smaller number of equivalent ones. Then, we derive the server's optimal reward $r_j^*(\boldsymbol{d})$ for any given data size $\boldsymbol{d}$ in the contract (Lemma \ref{reward}). Finally, we calculate   the optimal contract $\boldsymbol{\phi}^*$ (Theorem \ref{thm4}).

We denote the  set of user types incentivized by the sever as $\mathcal{J}'\triangleq\{1_{\mathcal{J}'},2_{\mathcal{J}'},...,J'_{\mathcal{J}'}\}$ (to be derived in Stage II), in which we reindex the types   according to the ascending order of marginal cost $\theta$.\footnote{For example,  if types $1$, $3$, and  $5$ are in this set, then we reindex them by $1_{\mathcal{J}'}\triangleq 1$, $2_{\mathcal{J}'}\triangleq 3$, and $3_{\mathcal{J}'}\triangleq 5$.}
%Lemma \ref{lm1} characterizes the contract feasibility:
%\vspace{-0.5mm}
%\begin{lemma}
%	\label{lm1}
%	Given an incentivized user type set $\mathcal{J}'$, a contract $ \boldsymbol{\phi}$  is feasible if and only if:
%	\begin{enumerate}
%		\item[a)] for  user types in $\mathcal{J}'$, the contract items satisfy the following three conditions:
%		\vspace{-1mm}
%		\begin{enumerate}
%			\item[a.1)] $r_{J'}-\theta_{J'}d_{J'}-c(\boldsymbol{p}) \ge 0$;
%			\item[a.2)] $ r_{1'}\ge ...\ge r_{J'} \ge 0$ and $ d_{1'}\ge ...\ge d_{J'}\ge 0$;
%			\item[a.3)] $r_{{j+1}}\hspace{-0.5mm}+\hspace{-0.5mm}\theta_{j}(d_{j}\hspace{-0.5mm}-\hspace{-0.5mm}d_{{j+1}})\hspace{-0.5mm} \le\hspace{-0.5mm} r_{j} \hspace{-0.5mm}\le\hspace{-0.5mm} r_{{j+1}}\hspace{-0.5mm}+\hspace{-0.5mm}\theta_{{j+1}}(d_{j}\hspace{-0.5mm}-\hspace{-0.5mm}d_{{j+1}}), j\hspace{-0.5mm}\in\hspace{-0.5mm} \mathcal{J}'\hspace{-0.5mm}.$
%		\end{enumerate}
%		\item[b)] for any user type $j\notin \mathcal{J}'$, $d_{j}=r_{j}=0$.
%	\end{enumerate}
%\end{lemma}
The following Lemma \ref{reward} characterizes the optimal rewards for any feasible data size:
\begin{lemma}
	\label{reward}
	For any given data size $\boldsymbol{d}=\{d_j\}_{j\in \mathcal{J}}$ (even if it is not optimal),  the  optimal rewards satisfy:
	\vspace{-0.5mm}
	\begin{itemize}
		\item 	for any user type $j \in  \mathcal{J}'$,
		\vspace{-1mm}
		\begin{equation*}
		\begin{split}
		&\hspace{-3mm}r_{j}^*(\boldsymbol{d},\boldsymbol{p})\hspace{-1mm}=\\& \hspace{-4mm} \left\{ \hspace{-2mm}
		\begin{array}{l}
		\hspace{-0.5mm}\theta_{j} d_{j}\hspace{-0.8mm}+\hspace{-0.6mm}c(\boldsymbol{p}), \hspace{38.4mm} \rm{if }\; $$j\hspace{-0.5mm}=\hspace{-0.5mm}{J}'_{\mathcal{J}'},$$\\
		\hspace{-0.5mm}\theta_{j} d_{j}\hspace{-0.8mm}+\hspace{-0.8mm}\sum_{k={(j+1)_{\hspace{-0.5mm}\mathcal{J}'}}}^{{J}'_{\hspace{-0.5mm}\mathcal{J}'}}\hspace{-0.5mm}(\theta_{k}\hspace{-1mm}-\hspace{-0.5mm}\theta_{{k-1}})d_{k}\hspace{-0.8mm}+\hspace{-0.8mm}c(\boldsymbol{p})\hspace{-0.5mm},   \rm{if }\; $$j\hspace{-0.5mm}=\hspace{-0.5mm}1'_{\hspace{-0.5mm}\mathcal{J}'}\hspace{-0.3mm},...,({J}'\hspace{-1.5mm}-\hspace{-1mm}1)_{\hspace{-0.5mm}\mathcal{J}'}\hspace{-0.5mm},$$\\
		\end{array} \right.
		\end{split}
		\end{equation*}
		\vspace{-2mm}
		\item for any user type $j\notin \mathcal{J}'$, $r_j^*=0$.
	\end{itemize}
	\vspace{-1mm}
\end{lemma}

Based on Lemma \ref{reward}, the following theorem characterizes the server’s optimal contract given any network price:
\begin{theorem}
	\label{thm4}
	Given the network operator's price $\boldsymbol{p}$, there exists a unique threshold type $x^*(\boldsymbol{p})$,
	\vspace{-1mm}
	\begin{equation}
	x^*(\boldsymbol{p})=\arg\min_{x\in\mathcal{J}}W_S(x,\boldsymbol{p}),
	\end{equation}
	\vspace{-4mm}
	
	\noindent where $W_S(x,\boldsymbol{p})$ is given in \eqref{wsp} on the next page, 
	%	\begin{equation}
	%	W_S(x,\boldsymbol{p})=\begin{cases}
	%	\left(\frac{2\xi}{I_x}\right)^{\frac{1}{3}}\left[\left(\sum_{j=1}^{x}I_j\right)\theta_x - \left(\sum_{j=1}^{x-1}I_j\right)\theta_{x-1}\right]^{\frac{1}{3}}\\ \hspace{31mm}+ \xi \left(\frac{1}{I_x^{\frac{1}{3}}(2\xi)^{\frac{2}{3}}} - \frac{\left(\sum_{j=1}^{x}I_j\right)\left(\sum_{j=1}^{x-1}I_j\right)(\theta_x-\theta_{x-1})d^{\max}}{I_x} + \sum_{j=1}^{x}I_jc(\boldsymbol{p}) \right),\\ 
	%	\hspace{16mm}\rm{if}\; $$d_j^*=d^{\max},\forall j<x$$\; \rm{and} \; $$\frac{I_x^{\frac{2}{3}}}{(2\xi)^{\frac{2}{3}}\left(\sum_{j=1}^{x}I_j\right)\left[\left(\sum_{j=1}^{x}I_j\right)\theta_x - \left(\sum_{j=1}^{x-1}I_j\right)\theta_{x-1}\right]^{\frac{2}{3}}}\le d^{\max},$$\\
	%	\frac{1}{\sqrt{\sum_{j=1}^xI_jd^{\max}}} +\xi \left(\sum_{j=1}^xI_j\theta_x d^{\max} +  \sum_{j=1}^{x}I_jc(\boldsymbol{p})\right),\\ 
	%	\hspace{16mm}\rm{if}\; $$d_j^*=d^{\max},\forall j<x$$\; \rm{and} \; $$\frac{I_x^{\frac{2}{3}}}{(2\xi)^{\frac{2}{3}}\left(\sum_{j=1}^{x}I_j\right)\left[\left(\sum_{j=1}^{x}I_j\right)\theta_x - \left(\sum_{j=1}^{x-1}I_j\right)\theta_{x-1}\right]^{\frac{2}{3}}}> d^{\max},$$\\
	%	\infty, \hspace{10mm}\rm{if}\; $$\exists  j<x, d_j^*<d^{\max}.$$
	%	\end{cases}
	%	\end{equation}
		\begin{figure*}
		\small
		\begin{equation}
		\label{wsp}
		\begin{split}
		W_S(x,\boldsymbol{p})\hspace{-1mm}=\hspace{-1mm}\begin{cases}
		\frac{1}{\sqrt{\sum_{j=1}^xI_jd^{\max}}} +\xi \left(\sum_{j=1}^xI_j\theta_x d^{\max} +  \sum_{j=1}^{x}I_jc(\boldsymbol{p})\right), \hspace{4.5mm}\rm{if}\; $$d_j^*=d^{\max},\forall j<x$$\; \rm{and} \; $$d^{\max}<d^{th1},$$\\
		\hspace{-0.5mm}\left(\frac{2\xi}{I_x}\right)^{\frac{1}{3}}\hspace{-1mm}\left[\hspace{-0.5mm}\left(\hspace{-0.5mm}\sum_{j=1}^{x}\hspace{-0.5mm}I_j\hspace{-0.5mm}\right)\hspace{-0.5mm}\theta_x\hspace{-1mm}-\hspace{-1mm} \left(\hspace{-0.5mm}\sum_{j=1}^{x-1}\hspace{-0.5mm}I_j\hspace{-0.5mm}\right)\hspace{-0.5mm}\theta_{x-1}\right]^{\frac{1}{3}}\hspace{-1.5mm}+\hspace{-1mm} \xi \left(\frac{1}{I_x^{\frac{1}{3}}(2\xi)^{\frac{2}{3}}} \hspace{-1mm}-\hspace{-1mm} \frac{\left(\sum_{j=1}^{x}I_j\right)\left(\sum_{j=1}^{x-1}I_j\right)(\theta_x - \theta_{x-1})d^{\max}}{I_x} \hspace{-0.5mm}+ \hspace{-0.5mm}\sum_{j=1}^{x}I_jc(\boldsymbol{p}) \right),\\ 
		\hspace{78mm}\rm{if}\; $$d_j^*=d^{\max},\forall j<x$$\; \rm{and} \; 
		$$d^{th1}\le d^{\max}<d^{th2},$$\\
		\infty,\; \hspace{72.5mm}\rm{if}\; $$\exists  j<x, d_j^*<d^{\max}$$\; \rm{or} \; $$d^{\max}\ge d^{th2}.$$
		\end{cases}
		\end{split}
		\vspace{-3mm}
		\end{equation}
		\normalsize
		\vspace{-6mm}
		\rule[0.15\baselineskip]{\textwidth}{0.03em}
	\end{figure*}
	such that  the server's optimal incentivized type set is $\mathcal{J}'^*\triangleq\{1,2,$ $...,	x^*(\boldsymbol{p})\}$ and the optimal contract item for  type-$j$ users is
	\vspace{-1mm}
	\begin{equation*}
	\begin{split}
	&\phi_j^*(\boldsymbol{p})=(d_j^\ast(\boldsymbol{p}), r_j^\ast(\boldsymbol{p})) = \\
	&\hspace{-0.5mm}\begin{cases}
	\hspace{-1.5mm}\left(\hspace{-0.5mm}d^{\max}\hspace{-0.5mm},\hspace{-0.5mm} \theta_{x^*\hspace{-0.5mm}(\hspace{-0.3mm}\boldsymbol{p}\hspace{-0.3mm})\hspace{-0.5mm}-\hspace{-0.5mm}1}d^{\max}\hspace{-0.5mm}+\hspace{-0.5mm}(\hspace{-0.5mm}\theta_{x^*\hspace{-0.5mm}(\hspace{-0.3mm}\boldsymbol{p}\hspace{-0.3mm})}\hspace{-1mm}-\hspace{-0.5mm}\theta_{{x^*\hspace{-0.5mm}(\hspace{-0.3mm}\boldsymbol{p}\hspace{-0.3mm})}\hspace{-0.5mm}-\hspace{-0.5mm}1}\hspace{-0.5mm}) d_{x^*\hspace{-0.5mm}(\hspace{-0.3mm}\boldsymbol{p}\hspace{-0.3mm})}^*\hspace{-1mm}+\hspace{-0.5mm}c(\boldsymbol{p})\hspace{-0.5mm}\right)\hspace{-1mm},\hspace{-1mm}\forall j\hspace{-0.5mm} <\hspace{-0.5mm}{x^*\hspace{-0.5mm}(\boldsymbol{p})},\\
	\hspace{-1.5mm}\left(d_{x^*(\boldsymbol{p})}^*, \theta_{x^*(\boldsymbol{p})} d_{x^*(\boldsymbol{p})}^*+c(\boldsymbol{p})\right),\hspace{26.2mm}j={x^*(\boldsymbol{p})},\\
	\hspace{-0.5mm}\boldsymbol{0},  \hspace{66.3mm}\forall j\hspace{-0.5mm} >\hspace{-0.5mm}{x^*\hspace{-0.5mm}(\boldsymbol{p})},
	\end{cases}
	\end{split}
	\end{equation*}
	\vspace{-3.5mm}
	
	\noindent where $d_{x^*(\boldsymbol{p})}^*$ is
	\vspace{-1mm}
%	\begin{figure*}
%		\small
		\begin{equation}
		\label{dx}
	%	d_{x^*(\boldsymbol{p})}^*\hspace{-1mm}=\hspace{-1mm}
		\begin{cases}		
		\hspace{-1mm}\frac{1}{I_{x^*\hspace{-0.5mm}(\boldsymbol{p})}^{\frac{1}{3}}(2\xi)^{\frac{2}{3}}\hspace{-0.5mm}\left[\left(\sum_{j=1}^{{x^*\hspace{-0.5mm}(\boldsymbol{p})}}I_j\right)\theta_{x^*\hspace{-0.5mm}(\boldsymbol{p})}\hspace{-0.5mm}-\hspace{-0.5mm} \left(\sum_{j=1}^{{x^*\hspace{-0.5mm}(\boldsymbol{p})}-1}I_j\right)\theta_{{x^*\hspace{-0.5mm}(\boldsymbol{p})}\hspace{-0.5mm}-1}\right]^{\frac{2}{3}}}\\
		\hspace{18.2mm}-\frac{\left(\sum_{j=1}^{{x^*(\boldsymbol{p})}-1}I_j\right)d^{\max}}{I_{x^*(\boldsymbol{p})}},\rm{if}\; $$ d^{\max}\ge d^{opt},$$\\	
		\hspace{-1mm}d^{\max},\; \hspace{40mm}\rm{if}\; $$ d^{\max}<d^{opt}.$$\\	
		\end{cases}
		\end{equation}

%		\normalsize
%		\vspace{-5mm}
%		\rule[-12pt]{\textwidth}{0.05em}
%	\end{figure*}
\end{theorem}
	Due to space limit, we do not show the complex expressions of thresholds $d^{th1}$, $d^{th2}$, and $d^{opt}$ in \eqref{wsp} and \eqref{dx} here.
	
%We  denote the set of users incentivized by the server as $\mathcal{X}_S$. Then, Theorem \ref{thm4} shows that $\mathcal{X}^*_S$ includes the users of  types $\{1,2,...,	x^*(\boldsymbol{p})\}$, i.e., users with relatively small marginal costs.
Theorem \ref{thm4} shows that it is optimal for the server to incentivize the users with relatively small  marginal costs.
The server   sets positive contract items for  the incentivized user types and zero for the not incentivized ones. 
Under such a contract, the threshold type users (i.e., type-$x^*(\boldsymbol{p})$ users) only obtain a zero payoff, as the server's optimal rewards just cover their training costs and network costs. Users with marginal costs smaller than type-$x^*(\boldsymbol{p})$ users  will obtain positive payoffs.
The specific values of rewards and data sizes depend on the network operator's pricing $\boldsymbol{p}$ in Stage I.
%In this case, the optimal network usage at equilibrium may not be water-filling and the optimal price at equilibrium  may not be the same. However, since users are congestion-sensitive, so they will reach the equilibrium voluntarily instead of under some instructions of the network operator.

%\subsection{Network Operator's Optimal Price}
%We denote the set of users selected by the server as $\mathcal{X}$, i.e., $\mathcal{X}$ contains type-$1$ to type-$x$ users.
\vspace{-1.5mm}
\subsection{Network Operator's Optimal Pricing in Stage I}
\vspace{-0.5mm}
\label{networko}
Considering the server's optimal contract in Stage II and users' optimal strategies in Stage III, the network operator needs to design  prices $\boldsymbol{p}\hspace{-0.5mm}\triangleq\hspace{-0.5mm}\{p(t)\}_{t\in \mathcal{T}}$ to maximize its profit:

\vspace{-1mm}
\begin{tcolorbox}
	\begin{problem}[Network Operator's Pricing  in Stage I]
		\label{q1v}
		\vspace{-1mm}
		\begin{equation*}
		\begin{split}
		\max \quad&\sum_{i\in \mathcal{I}}p(t_i^*(\boldsymbol{p}))-\gamma \sum_{t\in \mathcal{T}} \bigg(\sum_{i\in \mathcal{I}}\mathds{1}_{t_i^*(\boldsymbol{p})=t}+h(t)\bigg)^2\\
		\rm{s.t.}\quad & 0\le p(t)\le p_0,\; t\in \mathcal{T}\\
		%& t_i={\arg\min}_{t\in \mathcal{T}} p(t)\hspace{-1mm} +\hspace{-1mm}\beta \left(\sum_{k\in \mathcal{I}}\mathds{1}_{t_k=t}+h(t)\right)^2\hspace{-1.5mm}, \forall i\in \mathcal{X}_O\\
		\rm{var.}\quad & \{p(t)\}_{t\in \mathcal{T}}
		\end{split}
		\end{equation*}
	\end{problem}
\end{tcolorbox}
\vspace{-1mm}
It is challenging to solve Problem \ref{q1v} for several reasons. First, pricing  not only directly affects the network operator's revenue but also indirectly determines the network cost by influencing users' decisions. Second, the network operator needs to consider all users'  optimal  time choices, the complex form of which makes the optimization problem   non-convex. 

We tackle the above challenges by decomposing the analysis of Problem \ref{q1v} into two steps. First, we compute the network operator's optimal  network  demand distribution (i.e., the number of users  in each time slot) under a given set of participating users (Lemma \ref{lm6}), by leveraging proper transformations of variables and functions. Given this demand distribution, we then compute the network operator's optimal prices in Theorem \ref{thmno}, by decomposing the multi-variable optimization into sequential single-variable optimizations.  
%by manipulating each user's payoff through pricing. Different load distributions will lead to different network costs of network operator. 

For the convenience of presentation, we first introduce several notations. We denote the set of users selected by the network operator (i.e., those can afford the network prices of the chosen time slots and participate in the federated learning) by $\mathcal{X}_O$ (to be derived in Theorem \ref{thmno}). We denote $n_t$ as the number of users   in this federated learning system  who upload results at time slot $t$. We define the set of time slots that will be chosen by at least one user as $\mathcal{Q}$, and the set of time slots that will not be chosen by any user as $\bar{\mathcal{Q}}$, i.e., $\mathcal{Q}\cup\bar{\mathcal{Q}}=\mathcal{T}$.
In other words, for each time slot $t\in \mathcal{Q}$, we have $n_t>0$; and for each time slot $t\in \bar{\mathcal{Q}}$, we have $n_t=0$. 
\vspace{-0.5mm}
\begin{lemma}
	\label{lm6}
	Given a set of selected users $\mathcal{X}_O$, the network operator's optimal time slot sets $\mathcal{Q}^*(\mathcal{X}_O)$ and $\bar{\mathcal{Q}}^*(\mathcal{X}_O)$ are
	%\vspace{-1mm}
	\begin{equation}
	\begin{split}
	\mathcal{Q}^*(\mathcal{X}_O)\hspace{-0.3mm}=\hspace{-0.3mm}\{t:h(t)(\beta h(t)+2\gamma)\hspace{-0.3mm}\le\hspace{-0.3mm} h(\tilde{t})(\beta h(\tilde{t})+2\gamma)\},\\
	\bar{\mathcal{Q}}^*(\mathcal{X}_O)\hspace{-0.3mm}=\hspace{-0.3mm}\{t:h(t)(\beta h(t)+2\gamma)\hspace{-0.3mm}> \hspace{-0.3mm}h(\tilde{t})(\beta h(\tilde{t})+2\gamma)\},\\
	\end{split}
	\end{equation}
	where the threshold time slot $\tilde{t}$ is the unique value that makes $	\mathcal{Q}^*(\mathcal{X}_O)$ and $\bar{\mathcal{Q}}^*(\mathcal{X}_O)$ satisfy
	\vspace{-1mm}
	\begin{subnumcases}{\label{15}}
	\hspace{-1mm}\max_{t\hspace{-0.3mm}\in\hspace{-0.3mm} \mathcal{Q}^*\hspace{-0.5mm}(\hspace{-0.5mm}\mathcal{X}_O\hspace{-0.5mm})}\hspace{-1.5mm}h(t)(\beta h(t)\hspace{-0.5mm}+\hspace{-0.5mm}2\gamma)\hspace{-0.5mm}\le \hspace{-1mm}-\hspace{-0.5mm}\lambda\hspace{-0.5mm}\le \hspace{-1.5mm}\min_{t\hspace{-0.3mm}\in\hspace{-0.3mm} \bar{\mathcal{Q}}^*\hspace{-0.5mm}(\hspace{-0.5mm}\mathcal{X}_O\hspace{-0.5mm})}\hspace{-1.5mm}h(t)(\beta h(t)\hspace{-0.5mm}+\hspace{-0.5mm}2\gamma)\hspace{-0.5mm},\\
	\hspace{-1mm}\sum_{t\hspace{-0.3mm}\in\hspace{-0.3mm} \mathcal{Q}^*\hspace{-0.5mm}(\hspace{-0.5mm}\mathcal{X}_O\hspace{-0.5mm})}\hspace{-4mm}\sqrt{(\beta h(t)\hspace{-0.5mm}-\hspace{-0.5mm}\gamma)^2\hspace{-1mm}-\hspace{-0.5mm}3\beta \lambda}\hspace{-0.5mm}=\hspace{-0.5mm}3\beta |\mathcal{X}_O\hspace{-0.3mm}|\hspace{-0.5mm}+\hspace{-3mm}\sum_{t\hspace{-0.3mm}\in\hspace{-0.3mm} \mathcal{Q}^*\hspace{-0.5mm}(\hspace{-0.5mm}\mathcal{X}_O\hspace{-0.5mm})}\hspace{-3mm}(2\beta h(t)\hspace{-0.5mm}+\hspace{-0.5mm}\gamma)\hspace{-0.5mm}.
	\end{subnumcases}	
	The network operator's  optimal demand distribution is 
	\vspace{-1mm}
	\begin{equation*}
	\hspace{-72mm}n_t^* (\mathcal{X}_O)=
	\end{equation*}
	\vspace{-4.5mm}
	\begin{subnumcases}{\label{ho}}
	\hspace{-1mm}\frac{\sqrt{(\beta h(t)\hspace{-0.3mm}-\hspace{-0.3mm}\gamma)^2\hspace{-0.3mm}-\hspace{-0.3mm}3\beta \lambda}\hspace{-0.3mm}-\hspace{-0.3mm}(2\beta h(t)\hspace{-0.3mm}+\hspace{-0.3mm}\gamma)}{3\beta}, \forall t\hspace{-0.3mm}\in \hspace{-0.3mm}\mathcal{Q}^*(\mathcal{X}_O),\\
	0,\;\hspace{51mm}\forall t\hspace{-0.3mm}\in \hspace{-0.3mm}\bar{\mathcal{Q}}^*(\mathcal{X}_O).
	\end{subnumcases}
\end{lemma}
Lemma \ref{lm6} indicates that the network operator wants users to choose the time slots with   small  values of $h(t)(\beta h(t)+2\gamma)$, which   can be manipulated by the network operator though  proper prices (to be shown in Theorem \ref{thmno}). The  criterion $h(t)(\beta h(t)+2\gamma)$ indicates that the network operator  considers the network costs of both itself (indicated by term $2\gamma h(t)$) and users (indicated by term $\beta h(t)^2$). 

Moreover, Lemma \ref{lm6} shows that the time slots with less background network demands encourage the federated learning users' selection  but still have less total network demands ($n_t^*+h(t)$). This is because the network operator needs to consider users' total congestion cost in each selected time slot, which  cubically increases in the number of users who choose that slot. We will illustrate this by simulations in Section \ref{sim}.

Based on  the optimal demand distribution in Lemma \ref{lm6}, we present the network operator's optimal pricing in Theorem \ref{thmno}.
\vspace{-0.5mm}
%\vspace{-1mm}
\begin{theorem}
	\label{thmno}
	The network operator's optimal selected user set 
	$\mathcal{X}_O^*$ contains users of types $\{1,2,...,x_O^*\}$ with 
	\vspace{-1.5mm}
		\small
	\begin{equation}
	\label{14}
	\begin{split}
	\hspace{-1mm}x_O^*&=\arg \max_{x_O\in \mathcal{J}}\left(|\mathcal{X}_O|\tilde{C}^*(x_O)-\right.
	\\&\hspace{1.6mm} \beta\sum_{i\in \mathcal{X}_O} \left(n_{t_i}^*(\mathcal{X}_O)+h(t_i)\right)^2-\gamma \sum_{t\in \mathcal{T}} \left(n_t^*(\mathcal{X}_O)+h(t)\right)^2\Big),
	\end{split}
	\end{equation}
		\vspace{-3mm}
		\normalsize
		
\noindent	where
	\small
		\vspace{-1mm}
	\begin{equation*}
	\label{cx}
	\begin{split}
	&\tilde{C}^*\hspace{-0.5mm}(\hspace{-0.5mm}x_O\hspace{-0.5mm})\hspace{-0.7mm}=\hspace{-0.5mm}\max\hspace{-0.5mm}\Big\{\hspace{-0.5mm}c\hspace{-0.8mm}: \hspace{-0.5mm} c\hspace{-0.5mm}<\hspace{-0.5mm}p_0\hspace{-0.5mm}+\hspace{-1.5mm}\min_{t\in \bar{\mathcal{Q}}^*\hspace{-0.5mm}(\hspace{-0.5mm}\mathcal{X}_O\hspace{-0.5mm})}\hspace{-1mm}\beta h(t)^2\hspace{-0.5mm},\hspace{-0.5mm} W_S\hspace{-0.2mm}(\hspace{-0.2mm}x_O\hspace{-0.2mm},\hspace{-0.2mm}\boldsymbol{p})\hspace{-0.7mm}\le\hspace{-0.5mm}\min_{j\in \mathcal{J}}\hspace{-0.2mm}W_S\hspace{-0.2mm}(j,\boldsymbol{p})\hspace{-0.2mm}, \\
	& \max_{t\hspace{-0.2mm}\in\hspace{-0.2mm} \mathcal{Q}^*\hspace{-0.5mm}(\hspace{-0.5mm}\mathcal{X}_O\hspace{-0.5mm})}\hspace{-1mm}\left(\hspace{-0.5mm}\beta \left(n_t^*\hspace{-0.5mm}(\hspace{-0.5mm}\mathcal{X}_O\hspace{-0.5mm})\hspace{-0.5mm}+\hspace{-0.5mm}h(t)\right)^2\hspace{-0.5mm}\right)\hspace{-0.9mm}\le\hspace{-0.7mm} c \hspace{-0.7mm}\le \hspace{-0.5mm} p_0\hspace{-0.7mm}+\hspace{-1mm}\min_{t\hspace{-0.2mm}\in\hspace{-0.2mm} \mathcal{Q}^*\hspace{-0.5mm}(\hspace{-0.5mm}\mathcal{X}_O\hspace{-0.5mm})}\hspace{-1mm}\left(\beta\hspace{-0.5mm} \left(n_t^*\hspace{-0.5mm}(\hspace{-0.5mm}\mathcal{X}_O\hspace{-0.5mm})\hspace{-0.5mm}+\hspace{-0.5mm}h(t)\right)^2\hspace{-0.5mm}\right) \hspace{-1mm}\Big\}\hspace{-0.5mm}.
	\end{split}
	\end{equation*}
		\vspace{-4mm}
		\normalsize

\noindent	The network operator's optimal prices in Stage I are
%	\vspace{-1mm}
		\small
%	\begin{equation*}
%	\hspace{-75mm}p(t)^*=
%	\end{equation*}
	\vspace{-1mm}
	\begin{equation*}
	p(t)^*\hspace{-1.5mm}=\hspace{-1mm}
	\begin{cases}
	\hspace{-0.5mm}\tilde{C}^*\hspace{-0.4mm}(x_O^*) - \beta \left(n_{t}^*(\mathcal{X}_O^*)+h(t)\right)^2, \hspace{19.6mm}t\hspace{-0.8mm}\in\hspace{-0.8mm} \mathcal{Q}^*\hspace{-0.5mm}(\mathcal{X}_O^*),\\
	\hspace{-1.5mm}\text{any value} \hspace{-0.5mm}\in \hspace{-1mm}\left(\hspace{-0.5mm}\max_{t\in \bar{\mathcal{Q}}^*\hspace{-0.5mm}(\hspace{-0.5mm}\mathcal{X}_O^*\hspace{-0.5mm})}\hspace{-0.5mm}\{\tilde{C}\hspace{-0.2mm}(x_O^*\hspace{-0.5mm})\hspace{-0.5mm} -\hspace{-0.5mm} \beta h(t)^2\}\hspace{-0.3mm},p_0\right]\hspace{-1mm}, t\hspace{-0.8mm}\in\hspace{-0.8mm} \bar{\mathcal{Q}}^*\hspace{-0.5mm}(\mathcal{X}_O^*).
 	\end{cases}
	\end{equation*}
		\normalsize
	\vspace{-5mm}
%	
%	\noindent	$n_{t}^*(x_O^*)$ is given in Lemma \ref{lm6}, and $\tilde{C}^*(x)$ is given in \eqref{cx}. 
%	\begin{figure*}
		
%		\vspace{-5mm}
%		\rule[-12pt]{\textwidth}{0.05em}
%	\end{figure*}
\end{theorem}
\vspace{-1.5mm}

Theorem \ref{thmno} indicates that the network operator needs to consider  a trade-off   among the prices, the number of participating  users, and its network resource cost. If the network operator sets  larger prices,  the number of participating users will decrease but the network resource cost will also decrease. The optimal prices in Theorem \ref{thmno} maximize the network operator's profit under such a trade-off. %; If the network operator sets a smaller price, then the number of participating users will increase but the network resource cost will also increase.

%Moreover, as shown in Lemma \ref{lm6} and Theorem \ref{thmno}, the total  network usage demands   and prices  are different in   different time slots. However,  as we shall see in Section \ref{h1}, when users are congestion tolerant (i.e., ignoring congestion costs), the network demand at the equilibrium   has a water-filling form and optimal network prices for chosen time slots are the same. 

Next, we focus on the analysis of the two-stage game under the horizontal structure.

\vspace{-1mm}
\section{Incentive Mechanism and Network Pricing Design in Two-Stage Game  (Horizontal Structure)}
\vspace{-0.5mm}
\label{h2}
For the horizontal structure, the analysis for users in Stage II  is the same as that of Stage III under the vertical structure  in Section \ref{users}. Next, we present the analysis of the equilibrium strategies of the server  and the network operator   in Stage I.

We first define the server and the network operator's non-cooperative game and its equilibrium as follows:

\vspace{-1mm}
\begin{tcolorbox}
	\begin{game}[Stage I: Game of Server and Network Operator under  Horizontal Structure]%\quad \\
		%	\vspace{-4.5mm}
		\label{df3}
		The game between the server and the network operator in Stage I is
		\begin{itemize}
			\item Players: server and network operator.
%		\end{itemize}
%			\end{game}
%	\end{tcolorbox}
%\begin{tcolorbox}
%		\begin{itemize}					
			\item Strategy space: the sever designs the contract items $\boldsymbol{\phi}=\{\left(d_{j}, r_{j}\right)\}_{j\in \mathcal{J}}$, where $d_j \in [0,d^{\max}]$ and $r_j\in [0,+\infty)$, for each $j\in \mathcal{J}$. The network operator sets its prices $\boldsymbol{p}=\{p(t)\}_{t\in \mathcal{T}}$ at each time slot $t$, where $p(t) \in [0,p_0]$, for each $t\in \mathcal{T}$.
			\item Payoff function: the server  minimizes its cost
			\vspace{-1mm}
			\begin{equation}
			\label{200a}
			\begin{split}
			W_S(\boldsymbol{\phi};\boldsymbol{p})=\frac{1}{\sqrt{\sum_{j\in \mathcal{J}}I_jd_j}} +\xi \sum_{j\in \mathcal{J}}I_jr_j,
			\end{split}
			\end{equation}
			\vspace{-3mm}
			
			\noindent and the network operator maximizes its profit
			\vspace{-1.5mm}
			\begin{equation*}
			\label{200c}
			\begin{split}
			\hspace{-2mm}W_{O}(\boldsymbol{p};\hspace{-0.5mm}\boldsymbol{\phi})\hspace{-0.8mm}=\hspace{-1.5mm}\sum_{i\in \mathcal{I}}p(t_i^*\hspace{-0.2mm}(\boldsymbol{p}))\hspace{-0.7mm}-\hspace{-0.7mm}\gamma \hspace{-1mm}\sum_{t\in \mathcal{T}}\hspace{-0.9mm} \bigg(\hspace{-0.6mm}\sum_{i\in \mathcal{I}}\hspace{-0.5mm}\mathds{1}_{t_i^*(\boldsymbol{p})\hspace{-0.2mm}=t}\hspace{-0.5mm}+\hspace{-0.5mm}h(t)\hspace{-0.7mm}\bigg)^2\hspace{-1.3mm}.
			\end{split}
			\end{equation*}
		\end{itemize}
	\end{game}
	\vspace{-3.5mm}
%	\end{tcolorbox}
%		\vspace{-3mm}
%		\begin{tcolorbox}
	%\begin{tcolorbox}
	\begin{defn}[Equilibrium of Game \ref{df3}]
		\label{df4}
		The equilibrium of Game \ref{df3} is a  profile $(\boldsymbol{\phi}^*,\boldsymbol{p}^*)$,  such that the server and the network operator achieve their minimum cost or maximum profit assuming each other is following the equilibrium strategy:
		%neither the server nor the network operator has an incentive to deviate from their chosen strategies, after considering each other's strategies:
		\vspace{-1mm}
		\begin{equation*}
		W_S(\boldsymbol{\phi}^*\hspace{-0.3mm};\boldsymbol{p}^*\hspace{-0.5mm})\hspace{-0.5mm}\le\hspace{-0.5mm} W_S(\boldsymbol{\phi};\boldsymbol{p}^*\hspace{-0.5mm}), \hspace{-0.5mm} \forall d_j \hspace{-1mm}\in\hspace{-1mm} [0,\hspace{-0.5mm}d^{\max}],\hspace{-0.5mm}r_j\hspace{-1mm}\in \hspace{-1mm}[0,\hspace{-0.5mm}+\infty),\hspace{-0.5mm} {j\hspace{-1mm}\in\hspace{-1mm} \mathcal{J}}\hspace{-0.8mm},
		\end{equation*}
		
		\vspace{-8mm}
		\begin{equation}
		\hspace{-0.1mm}W_{O}(\boldsymbol{p}^*;\boldsymbol{\phi}^*\hspace{-0.5mm}) \hspace{-0.5mm}\ge\hspace{-0.5mm} W_{O}(\boldsymbol{p};\boldsymbol{\phi}^*\hspace{-0.5mm}), \hspace{-0.5mm}\forall p(t) \hspace{-0.5mm}\in \hspace{-0.5mm}[0,p_0],  t\hspace{-0.5mm}\in\hspace{-0.5mm} \mathcal{T}.\hspace{0.2mm}
		\end{equation}
%		\begin{equation}
%		\begin{split}
%		&W_S(\boldsymbol{\phi}^*;\boldsymbol{p}^*\hspace{-0.5mm})\hspace{-0.5mm}\le\hspace{-0.5mm} W_S(\boldsymbol{\phi};\boldsymbol{p}^*\hspace{-0.5mm}), \hspace{-0.5mm} \forall \boldsymbol{\phi}\hspace{-0.5mm} \in\hspace{-0.5mm} ([0,d^{\max}]^I\hspace{-0.5mm}, [0,+\infty\hspace{-0.5mm})^I);\\% d_j \in [0,d^{\max}],r_j\in [0,+\infty), {j\in \mathcal{J}};\\
%		&W_{O}(\boldsymbol{p}^*;\boldsymbol{\phi}^*\hspace{-0.5mm}) \hspace{-0.5mm}\ge\hspace{-0.5mm} W_{O}(\boldsymbol{p};\boldsymbol{\phi}^*\hspace{-0.5mm}), \hspace{-0.5mm}\forall \boldsymbol{p}\hspace{-0.5mm} \in\hspace{-0.5mm} [0,p_0]^I. %p(t) \in [0,p_0], t\in \mathcal{T}.
%		\end{split}
%		\end{equation}
	\end{defn} 
	%\end{tcolorbox}
\end{tcolorbox}
\vspace{-1mm}
According to the definitions, we will first analyze the best responses of the server and network operator in Section \ref{brno}, and then find out the fixed point of the  best responses  (which is the equilibrium)   in Section \ref{combine}. 
\vspace{-1mm}
\subsection{Best Responses of Sever and Network Operator  in Game \ref{df3}}
\vspace{-0.5mm}
\label{brno}
First, the server's optimal contract $\boldsymbol{\phi}^*(\boldsymbol{p})$ given the network operator's prices $\boldsymbol{p}$ (i.e.,   best response of the  server) is the same as the analysis in Section \ref{servers}. 

Next, we present the network operator's optimal pricing $\boldsymbol{p}^*(\boldsymbol{\phi})$ given any server's contract $\boldsymbol{\phi}$ (i.e., best response of the network operator) in Lemma \ref{net}.
The main analysis difference from that under the vertical structure in Section \ref{networko} is that, the network operator does not know the server's optimal strategies under the horizontal structure here.  
\vspace{-0.5mm}
\begin{lemma}
	\label{net}
	Given the server's contract $\boldsymbol{\phi}$, the network operator's optimal selected user set $\mathcal{X}_O^*(\boldsymbol{\phi})$ is
	\vspace{-2mm}
	\small
	\begin{equation}
	\begin{split}
	\hspace{-0.6mm}\mathcal{X}_O^*(\boldsymbol{\phi})&=\arg \max_{\mathcal{X}_O\subseteq \mathcal{I}} \Big( |\mathcal{X}_O|\tilde{C}^*(\mathcal{X}_O,\boldsymbol{\phi})  -\\
	&   \beta\hspace{-1mm}\sum_{i\in \mathcal{X}_O}\hspace{-1mm} \left(n_{t_i}^*(\mathcal{X}_O)+h(t_i)\right)^2
	\hspace{-1mm}	-\hspace{-0.5mm}\gamma \sum_{t\in \mathcal{T}} \left(n_t^*(\mathcal{X}_O)+h(t)\right)^2\hspace{-0.5mm}\Big)\hspace{-0.5mm},
	\end{split}
	\end{equation}
	\vspace{-3mm}
	\normalsize
	
	\noindent where users' maximum acceptable  network cost $\tilde{C}^*(\mathcal{X},\boldsymbol{\phi})$ is
	%\begin{figure*}
	\small
	\vspace{-1.5mm}
	\begin{equation}
	\label{c}
	\begin{split}
	\hspace{-1.8mm}\tilde{C}^*\hspace{-0.5mm}(\mathcal{X}_O,\boldsymbol{\phi})\hspace{-0.5mm}=\hspace{-0.5mm}\max \left\{\hspace{-0.5mm}\tilde{C}\hspace{-0.5mm}:\hspace{-0.5mm} \max_{t\in \mathcal{Q}^*\hspace{-0.5mm}(\hspace{-0.5mm}\mathcal{X}_O\hspace{-0.5mm})}\hspace{-0.5mm}\big(\hspace{-0.5mm}\beta \left(n_t^*\hspace{-0.5mm}(\mathcal{X}_O\hspace{-0.5mm})\hspace{-0.5mm}+\hspace{-0.5mm}h(t)\right)^2\hspace{-0.5mm}\big)\hspace{-0.5mm}\le \hspace{-0.5mm}\tilde{C} \hspace{-0.5mm} \le\right.&\\
	p_0\hspace{-1mm}+ \hspace{-1.5mm}\min_{t\in \mathcal{Q}^*\hspace{-0.5mm}(\hspace{-0.5mm}\mathcal{X}_O\hspace{-0.5mm})}\hspace{-0.5mm}\big(\hspace{-0.5mm}\beta \left(n_t^*\hspace{-0.5mm}(\mathcal{X}_O\hspace{-0.5mm})\hspace{-0.5mm}+\hspace{-0.5mm}h(t)\right)^2\hspace{-0.5mm}\big)\hspace{-0.5mm}, \hspace{-0.5mm} \tilde{C}\hspace{-0.5mm}<\hspace{-0.5mm}p_0\hspace{-1mm}+\hspace{-1.5mm}\min_{t\in \bar{\mathcal{Q}}^*(\mathcal{X}_O)}\hspace{-0.5mm}\beta h(t)^2\hspace{-0.5mm},&\\
	\max\{r_i-\theta_i d_i\}_{i\notin \mathcal{X}_O}<\tilde{C}\le \min  \{r_i-\theta_i d_i\}_{i\in \mathcal{X}_O}\bigg\}.&
	\end{split}
	\end{equation}
		\vspace{-3mm}
	\normalsize
	
	\noindent The network operator's optimal prices are
			\small
	\vspace{-1mm}
	\begin{equation*}
\begin{split}
			&\hspace{-1mm}p(t)^*(\boldsymbol{\phi})=\\
&\hspace{-3mm}	\begin{cases}
\hspace{-1mm}	\tilde{C}^*(\mathcal{X}_O^*,\boldsymbol{\phi}) - \beta \left(n_{t}^*(\mathcal{X}_O^*)+h(t)\right)^2,\hspace{14.7mm} t\hspace{-0.7mm}\in\hspace{-0.7mm} \mathcal{Q}^*\hspace{-0.5mm}(\hspace{-0.3mm}\mathcal{X}_O^*\hspace{-0.3mm})\hspace{-0.1mm},\\
\hspace{-1.5mm}	\text{any value} \hspace{-0.7mm}\in\hspace{-1mm} \left(\hspace{-0.5mm}\max_{t\in \bar{\mathcal{Q}}^*}\hspace{-0.5mm}\{\hspace{-0.3mm}\tilde{C}^*\hspace{-0.5mm}(\hspace{-0.3mm}\mathcal{X}_O^*,\hspace{-0.3mm}\boldsymbol{\phi}\hspace{-0.3mm})\hspace{-1mm} - \hspace{-1mm}\beta h(t)^2\hspace{-0.3mm}\},p_0\hspace{-0.3mm}\right]\hspace{-1mm}, t\hspace{-0.7mm}\in\hspace{-0.7mm} \bar{\mathcal{Q}}^*\hspace{-0.5mm}(\hspace{-0.3mm}\mathcal{X}_O^*\hspace{-0.3mm})\hspace{-0.1mm}.
	\end{cases}
\end{split}
		\end{equation*}
			\normalsize
\end{lemma}
%under the horizontal structure, the network operator will set different prices and target at different sets of users from that under the vertical structure. This is mainly because the network operator does not know the server's optimal strategies under the horizontal structure, but it knows through backward induction under the vertical structure.  
%\begin{remark}
%	The complexity of Algorithm \ref{alg:B} is $\mathcal{O}(JT)$, where $J$ is number of user types and $T$ is the number time slots.
%\end{remark}
\vspace{-0.5mm}
Next, we combine the best responses of  the server and the network operator to obtain the equilibrium under the horizontal structure.

\vspace{-1mm}
\subsection{Equilibrium in Stage I and Structure Comparison}
\vspace{-0.5mm}
\label{combine}
For the convenience of presentation, we first introduce the following definition: %\footnote{The notations in \eqref{26} are consistent with Algorithm \ref{alg:B}.% if wants to consistent with paper, then substitute x^* with  $\mathcal{X}_O^*$}:
\vspace{-1mm}
\small
%\begin{equation}
%\label{24}
%\begin{split}
%x^*=\arg \max_{x\in \mathcal{J}}\Big( |\mathcal{X}|\tilde{C}^*(x) -\beta\sum_{i\in \mathcal{X}} \hspace{-0.5mm}\left(n_{t_i}^*(x)\hspace{-0.5mm}+\hspace{-0.5mm}h(t_i)\right)^2\hspace{-1mm}-\\\hspace{-1mm}\gamma\hspace{-0.5mm} \sum_{t\in \mathcal{T}} \left(n_t^*(x)\hspace{-0.5mm}+\hspace{-0.5mm}h(t)\right)^2\Big).
%\end{split}
%\end{equation}
\begin{equation}
\label{26}
\begin{split}
&H\triangleq\max\Big\{c(\boldsymbol{p}):W_S(x^*,\boldsymbol{p})\le\min_{j\in \mathcal{J}}W_S(j,\boldsymbol{p})\Big\}-\\ &
\max \hspace{-0.5mm}\Big\{ \hspace{-0.5mm}p_0\hspace{-0.5mm}+\hspace{-1.5mm}\min_{t\in \bar{\mathcal{Q}}^* \hspace{-0.5mm}(\hspace{-0.5mm}\mathcal{X}^* \hspace{-0.5mm})}\hspace{-0.5mm}\beta h(t)^2\hspace{-0.5mm}, p_0\hspace{-0.5mm}+\hspace{-1.5mm} \min_{t\in \mathcal{Q}^* \hspace{-0.5mm}( \hspace{-0.5mm}\mathcal{X}^* \hspace{-0.5mm})}\hspace{-0.5mm}\beta \left(n_t^*(\mathcal{X}^*\hspace{-0.5mm})\hspace{-0.5mm}+\hspace{-0.5mm}h(t)\right)^2\hspace{-0.5mm}\Big\},
\end{split}
\end{equation}
\normalsize
where $x^*$ equals $x_O^*$ in \eqref{14} if $\tilde{C}^*(x_O)$ in \eqref{14}  equals
%\begin{equation}
%\label{23}
%\begin{split}
%\tilde{C}^*\hspace{-0.5mm}(\hspace{-0.3mm}x\hspace{-0.3mm})\hspace{-0.8mm}=\hspace{-0.8mm}\max\hspace{-0.5mm} \Big\{\hspace{-0.8mm}\tilde{C}\hspace{-0.5mm}:\hspace{-0.5mm} \max_{t\in \mathcal{Q}^*}\hspace{-0.5mm}\big(\hspace{-0.3mm}\beta\hspace{-0.5mm} \left(n_t^*\hspace{-0.3mm}(\hspace{-0.3mm}x\hspace{-0.3mm})\hspace{-0.5mm}+\hspace{-0.5mm}h(t)\hspace{-0.3mm}\right)^2\hspace{-0.5mm}\big)\hspace{-1mm}\le\hspace{-0.5mm} \tilde{C} \hspace{-0.5mm}\le\hspace{-0.5mm} p_0\hspace{-0.5mm}+\\\hspace{-0.5mm} \min_{t\hspace{-0.3mm}\in\hspace{-0.3mm} \mathcal{Q}^*}\hspace{-1mm}\big(\hspace{-0.5mm}\beta\hspace{-0.5mm} \left(n_t^*\hspace{-0.5mm}(\hspace{-0.5mm}x\hspace{-0.5mm})\hspace{-1mm}+\hspace{-1mm}h(t)\hspace{-0.5mm}\right)^2\hspace{-0.5mm}\big),\hspace{-0.5mm}\tilde{C} \hspace{-1mm}<\hspace{-0.5mm}p_0\hspace{-1mm}+\hspace{-0.5mm}\min_{t\in \bar{\mathcal{Q}}^*}\hspace{-0.5mm}\beta h(t)^2\hspace{-0.5mm} \hspace{-0.5mm}\Big\}.
%\end{split}
%\end{equation}
\small
	\begin{equation}
\begin{split}
&\max\hspace{-0.5mm}\Big\{\hspace{-0.5mm}c\hspace{-0.8mm}: \hspace{-0.5mm} c\hspace{-0.5mm}<\hspace{-0.5mm}p_0\hspace{-0.5mm}+\hspace{-1.5mm}\min_{t\in \bar{\mathcal{Q}}^*\hspace{-0.5mm}(\hspace{-0.5mm}\mathcal{X}_O\hspace{-0.5mm})}\hspace{-1mm}\beta h(t)^2\hspace{-0.5mm},  \max_{t\hspace{-0.2mm}\in\hspace{-0.2mm} \mathcal{Q}^*\hspace{-0.5mm}(\hspace{-0.5mm}\mathcal{X}_O\hspace{-0.5mm})}\hspace{-1mm}\left(\hspace{-0.5mm}\beta \left(n_t^*\hspace{-0.5mm}(\hspace{-0.5mm}\mathcal{X}_O\hspace{-0.5mm})\hspace{-0.5mm}+\hspace{-0.5mm}h(t)\right)^2\hspace{-0.5mm}\right)\\
&\hspace{25mm}\le\hspace{-0.7mm} c \hspace{-0.7mm}\le \hspace{-0.5mm} p_0\hspace{-0.7mm}+\hspace{-1mm}\min_{t\hspace{-0.2mm}\in\hspace{-0.2mm} \mathcal{Q}^*\hspace{-0.5mm}(\hspace{-0.5mm}\mathcal{X}_O\hspace{-0.5mm})}\hspace{-1mm}\left(\beta\hspace{-0.5mm} \left(n_t^*\hspace{-0.5mm}(\hspace{-0.5mm}\mathcal{X}_O\hspace{-0.5mm})\hspace{-0.5mm}+\hspace{-0.5mm}h(t)\right)^2\hspace{-0.5mm}\right) \hspace{-1mm}\Big\}\hspace{-0.5mm},
\end{split}
\end{equation}
\normalsize
and $\mathcal{X}^*$ contains users of types $\{1,2,...,x^*\}$.

Given the definition of $H$, we present the equilibrium existence and equilibrium strategies  under the horizontal structure in Theorem \ref{coe}:
\begin{theorem}
	\label{coe}
	 If $H\ge 0$, the equilibrium  exists under the horizontal structure  and  is the same as that under the vertical structure. If $H<0$, the equilibrium does not exist under the  horizontal structure .
\end{theorem}
\vspace{-0.5mm}
The condition $H\ge 0$ in Theorem \ref{coe} means  that  under the horizontal structure,  both the network operator and the server obtain  the maximum  profit (or minimum cost) by incentivizing the same group of users. 
%Why is the same? because there is no double marginalization effect (i.e., increase the price level by level. the server does not pay the network operator, it is the users pay the network operator)
The equilibrium does not exist if  the network operator and the server are interested in incentivizing  different groups of users (i.e., $H<0$). % i.e., $\mathcal{X}_S^*=\mathcal{X}_O^*$.

Moreover, Theorem \ref{coe} shows that  the equilibrium under the horizontal structure (if it exists) is the same as the vertical one. %This is counter-intuitive, as we may presume that different interaction structures and games will lead to different  results. 
Two aspects contribute to such a phenomenon. First, under the vertical   structure,   it is  the users who choose the time slots and pay the network operator, which is the same as the interaction under the horizontal structure. Second, the equilibrium under horizontal structure only exists when the network operator and the sever coincidentally incentivize the same group of users, same as the decision alignment under the vertical structure.  %Therefore, the server and the network operator will not increase the price level by level (i.e., no double marginalization effect), which makes the equilibrium the same as that under the horizontal structure.%contract(si,ti,ri), contract design, which is another problem. 

%Third,  Corollary \ref{coe} indicates that the vertical structure is better than the  horizontal one. The horizontal structure does not always have an equilibrium and even if it does, the equilibrium is just the same as that in vertical structure. This is because  sequential decision making (under the vertical structure) can avoid that the  network operator and the server wants to incentivize different groups of users, but  simultaneous decision making (under the horizontal structure) cannot avoid it as they cannot observe  each other's behavior before making decisions.
To conclude, Theorem  \ref{coe} shows that  the vertical structure is better than the horizontal one, as the vertical structure always ensures the existence of an equilibrium which is no worse than that of the horizontal structure  (if it exists). %This implies that the horizontal structure can introduce more intense competition than the vertical one, which harms everyone's interests.%This implies that the network operator and server in the federated learning system should make sequential decision making (i.e., vertical structure) instead of simultaneous one (i.e., horizontal structure).

The analysis so far focused on federated learning applications with congestion-sensitive users. Next, we study the special scenario with congestion-tolerant users.

\vspace{-1mm}
\section{Congestion-Tolerant Users}
\vspace{-0.5mm}
%\label{Insights}
%%In this section, we will first compare the performance of the horizontal and vertical interaction structures for users,  server, and  network operator in Section \ref{compare}. Then, we study the case congestion-tolerant users in Section \ref{h1}.
%
%\vspace{-1mm}
%\subsection{Congestion-Tolerant Users}
%\vspace{-0.5mm}
\label{h1}
In this section, we study the special case where users are tolerant of congestion, i.e., users do not have congestion costs ($\beta=0$). This is motivated by some practical scenarios where users do not have high requirements on  network qualities\footnote{For example, compared with the users in automatic driving, mobile phone users participating in the next-word-prediction learning task usually care less about time delay. Mobile users themselves  may even intentionally delay the parameter uploading due to considerations such as battery conditions \cite{mcmahan2016communication}.}. 

The analysis of the two structures in Sections \ref{v2} and \ref{h2} is still applicable to this special case. Meanwhile, we will be able to reveal some additional new insights, as presented  in Propositions \ref{sameprice} and \ref{lma1}. %, and  \ref{soe}.
\vspace{-0.5mm}
\begin{proposition}
	\label{sameprice}
	When $\beta=0$,  at the equilibrium, each user $i$ chooses a time slot  with the  lowest price, i.e.,
	\vspace{-1mm}
	\begin{equation}
	t_i^*={\arg\min}_{t\in \mathcal{T}} p(t), \; \forall i\in \mathcal{I}.
	\end{equation}
\end{proposition}
\vspace{-0.5mm}
As users are network congestion tolerant, they only care about the network price when selecting the time slots.  However, if too many users choose the same time slot, the network operator’s  resource   cost at this time slot will be very large, which may increase the network price for this slot. Therefore, users’ network usage at the equilibrium depends on  the network operator’s  pricing. 

\begin{figure}
	\centering
	\includegraphics[width=1.9 in]{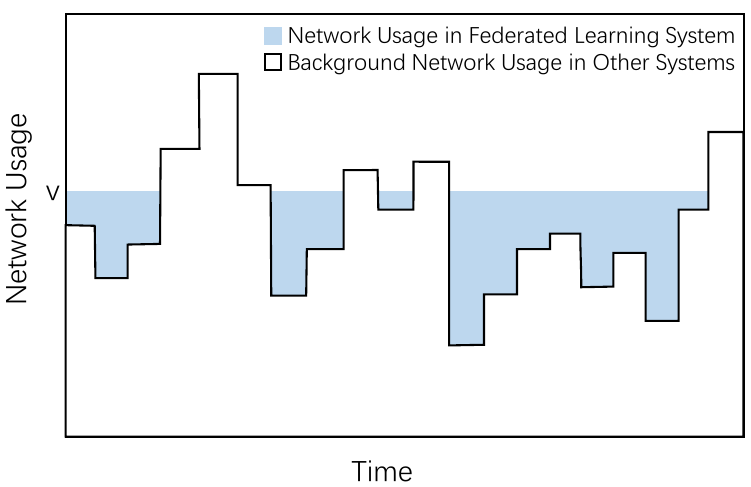}
	\vspace{-3.5mm}
	\caption{Illustration example of water-filling network usage distribution.}
	\vspace{-6mm}
	\label{water}
\end{figure}
\vspace{-0.5mm}
\begin{proposition}
	\label{lma1}
	When $\beta=0$, it is optimal for the network operator to have a water-filling network usage distribution and the same price for users' chosen   time slots. The total network usage of the chosen time slots (i.e., water level) $v$ satisfies
	\vspace{-1.5mm}
	\begin{equation}
	\sum_{t\in \mathcal{T}}[v-h(t)]^+=\sum_{t\in \mathcal{T}}  \bigg( \sum_{i\in \mathcal{I}} \mathds{1}_{t_i^*=t}\bigg),
	\end{equation}
	\vspace{-3.5mm}
	
	\noindent where 
		\vspace{-2mm}
	\begin{equation}
	[v-h(t)]^+\triangleq
	\begin{cases}
	v-h(t),\;\rm{if}\; $$v\ge h(t)$$,\\
	0,\hspace{11.5mm}\rm{if}\; $$v< h(t)$$.
	\end{cases}
	\end{equation}
\end{proposition}
\vspace{-0.5mm}
The water-filling network usage distribution is illustrated by the example in Fig.~\ref{water}, where participating users will choose the time slots with small background network usage such that the total network  usage in all chosen time slots will be the same (i.e., $v$).
When users are not concerned about their congestion costs, the network operator only needs to consider its own network cost in different time slots when designing the prices. %We can prove  that a water-filling network usage distribution  maximizes the its profit, and the optimal equilibrium prices are the same for all chosen time slots. 

Next, we use real-world datasets to validate the performance of our proposed mechanisms.

\vspace{-1.5mm}
\section{Simulation}
\vspace{-1mm}
\label{sim}
In this section, we perform numerical experiments to validate our analytical results and evaluate the performance of the proposed mechanisms. We  first introduce the experiment setting in Section \ref{setting}, then show the experiment results of the optimal contract and pricing in Section \ref{re}, and finally compare the performance of our mechanism with  two state-of-the-art benchmarks in Section \ref{ben}.
\vspace{-1.5mm}
\subsection{Experiment Setting}
\vspace{-1mm}
\label{setting}
%The dataset covers the Helsinki Metropolitan Area (HMA) in Finland.

\begin{figure*}
	\centering
	\begin{minipage}[t]{0.3\linewidth}
		\centering
		\includegraphics[width=2.2 in]{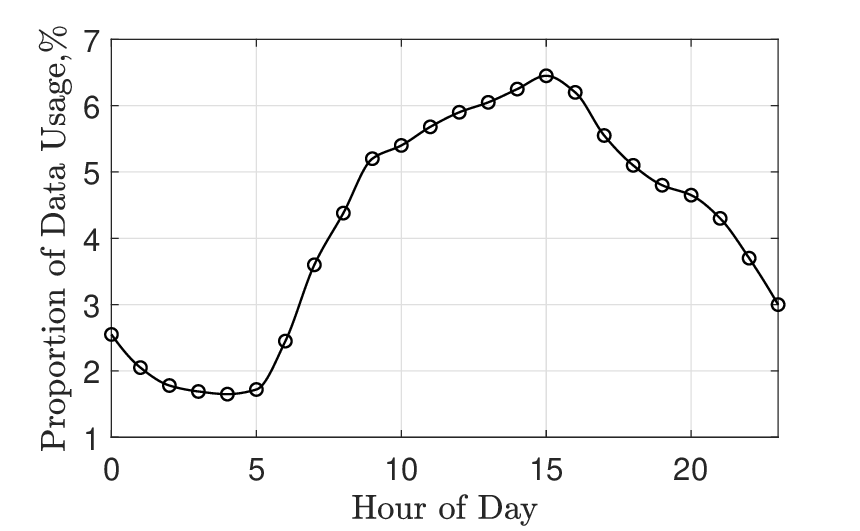}
		\vspace{-7mm}
		\caption{Hourly Mobile Phone Data Usage.}
		\vspace{-5mm}
		\label{ht}
	\end{minipage}
	\hspace{0.5mm}
	\begin{minipage}[t]{0.33\linewidth}
		\centering
		\includegraphics[width=2.2 in]{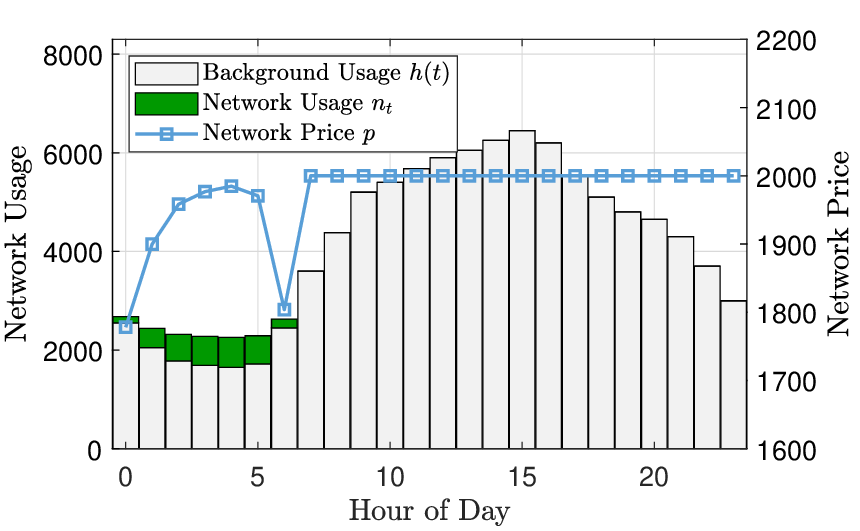}
		\vspace{-4mm}
		\caption{Network usage distribution and network operator's optimal prices in different time slots.}
		\vspace{-5mm}
		\label{price}
	\end{minipage}
\hspace{0.5mm}
\begin{minipage}[t]{0.33\linewidth}
	\centering
	\includegraphics[width=2.2 in]{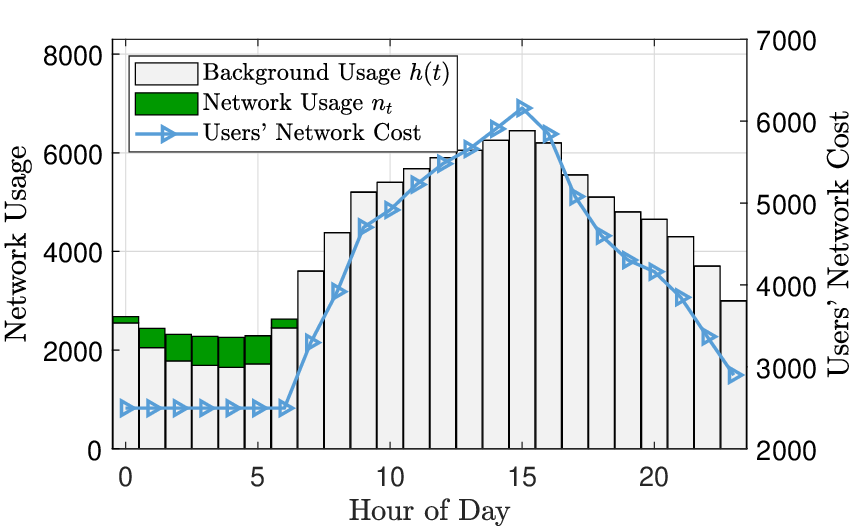}
	\vspace{-4mm}
	\caption{Network usage distribution and  users' network costs in different time slots.}
	\vspace{-5mm}
	\label{cost}
\end{minipage}
\end{figure*}

%The mobile phone data are allocated in statistical 250 m × 250 m grid cells using an advanced dasymetric interpolation method1 and validated against the population register data from Statistics Finland. Mobile phone data were provided by the largest mobile network operator in Finland.
We   use  the hourly mobile phone data usage  obtained from a real-world dataset as the background network usage distribution (as shown in Fig.~\ref{ht}). %The dataset  is provided by the Elisa Oyj, which is the largest mobile network operator in Finland. 
The dataset covers all base stations of the Elisa Oyj network operator in the Uusimaa region in Southern Finland, from late October 2017 till early January 2018 \cite{bergroth202224}. 

%\begin{figure}
%	\centering
%	\includegraphics[width=2.6 in]{ht.eps}
%	\vspace{-3mm}
%	\caption{Daily Mobile Phone Data Usage.}
%	\label{ht}
%\end{figure}

\begin{figure*}
	\centering
	\begin{minipage}[t]{0.32\linewidth}
		\centering
		\includegraphics[width=2.2 in]{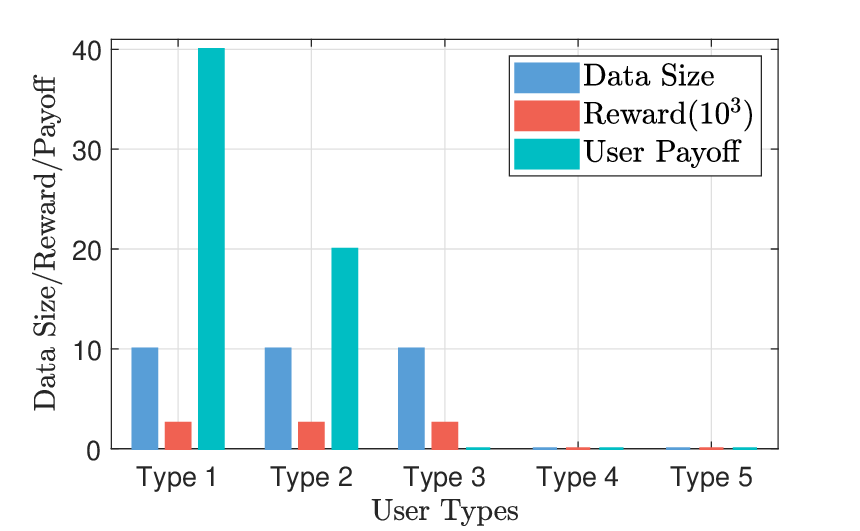}
		\vspace{-4mm}
		\caption{Server's optimal contract and users' payoffs for different types of users.}
		\vspace{-6mm}
		\label{server}
	\end{minipage}
	\hspace{0.5mm}
	\begin{minipage}[t]{0.32\linewidth}
		\centering
		\includegraphics[width=2.2 in]{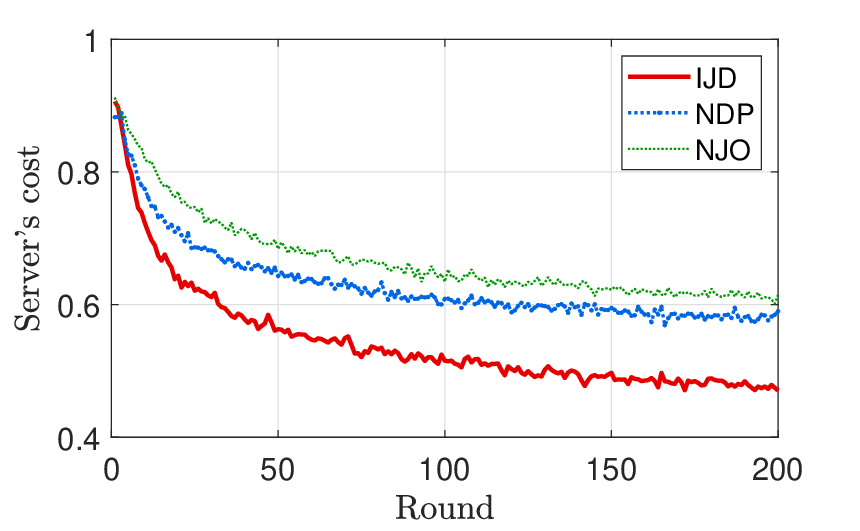}
		\vspace{-4mm}
		\caption{The comparison of server's costs under three mechanisms.}
		\vspace{-6mm}
		\label{sercost}
	\end{minipage}
	\hspace{0.5mm}
	\begin{minipage}[t]{0.32\linewidth}
		\centering
		\includegraphics[width=2.2 in]{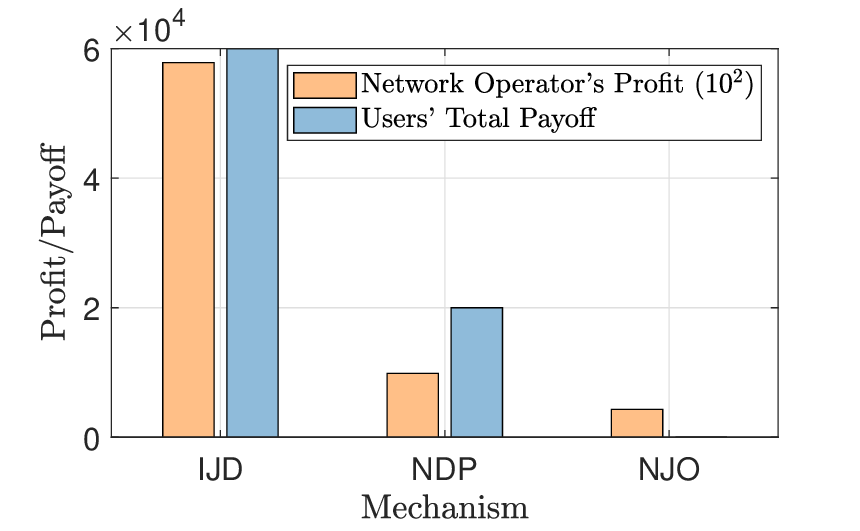}
		\vspace{-4mm}
		\caption{The comparison of network operator's profits and users' total payoffs  under three mechanisms.}
		\vspace{-6mm}
		\label{propay}
	\end{minipage}
\end{figure*}
Regarding the system parameters, we consider the time frame of one day that consists of 24 time slots, i.e., $T=24$. 
There are five types of users with marginal costs $\theta_1=2$, $\theta_2=4$, $\theta_3=6$, $\theta_4=8$, and $\theta_5=10$. Each type has $I_j=1000$ users, and the maximum data size that each user can contribute in one round is $d^{\max}=10$ MB. The maximum network price that the network operator can set is $p_0=2000$ cents. The  normalized total background network usage is $\sum_{t\in \mathcal{T}}h(t)=10^5$ users in other systems. We choose the congestion sensitive coefficients of users and the network operator as  $\beta=\gamma=10^{-4}$, and set the server's cost coefficient as $\xi=5\times10^{-10}$ to balance different parameters' units and orders of magnitude.

To obtain the experimental model accuracy loss, we consider that users with non-IID data   train a federated learning model on the CIFAR-10 dataset. Specifically, each user is randomly assigned 2 labels and each label has 50 data points. We assume that users' data distribution is independent of their marginal cost distribution. Our convolutional neural network (CNN) model consists of six $3\times3$ convolution layers (with 64, 64, 128, 128, 256, 256 channels, respectively, and every two  followed with $2\times 2$ max pooling), a Drop-out layer (0.5),  a fully-connected layer with 10 units  and ReLU activation, and  a final softmax output layer.

\vspace{-1mm}
\subsection{Experiment Results: Optimal Pricing and Contract}
\vspace{-1mm}
\label{re}
As the horizontal structure has no equilibrium or the same equilibrium with the vertical structure, next we only present the numerical results under the vertical structure\footnote{We can validate that under the same experiment setting in Section \ref{setting}, the equilibrium under the horizontal structure does not exist. If we  change the experiment setting, the equilibrium  under the horizontal structure can exist and will be the same as that under the vertical structure. Due to  space limit, we will not show the detailed simulation results under the horizontal structure.}.

%\begin{figure}
%	\centering
%	\includegraphics[width=2.6 in]{price1.eps}
%	\vspace{-3mm}
%	\caption{Network usage distribution and network operator's optimal prices at different time slots under the vertical structure.}
%	\label{price}
%\end{figure}
\subsubsection{Network Operator's Optimal  Demand Distribution}
Fig.~\ref{price} shows that when users are congestion-sensitive, the optimal network demand distribution is not a water-filling solution, i.e., the total network usages (green plus gray in Fig.~\ref{price})  in users' chosen time slots are not the same. 
Users will choose the time slots with small background network usages $h(t)$ (i.e., 0:00-6:00). Interestingly,  among these chosen time slots,  the  slots with smaller background usages (gray) still have smaller total network usages (green plus gray). 
This is consistent with the results in Lemma \ref{lm6}, because the network operator  does not want too many users choose the same time slot, as the total  congestion cost of users in a time slot cubically increases in the number of users who choose this time slot. %, even if this time slot has a small background network usage $h(t)$ (e.g., at 4:00). %If users are 
%This is because when the network operator designs the price in a time slot, he needs to consider users' total  congestion cost in this time slot, which quadratically increases in the number of users in this time slot (since users' total  congestion cost  quadratically increases in the number of users in this time slot,
\subsubsection{Network Operator's  Optimal Prices}
As shown in Fig.~\ref{price}, the optimal prices for   time slots chosen by at least one user are different, i.e., there is a larger price for a smaller total network usage. %This is to ensure a small total network cost (sum of the price and the congestion cost) for users. 
As shown in Fig.~\ref{cost}, under the optimal prices, users at different chosen time slots (i.e., 0:00-6:00) have the same minimum network cost (the sum of the price and the congestion cost), which is consistent with  Lemma  \ref{th4}.

\subsubsection{Server's  Optimal Contract}
As shown in Fig.~\ref{server}, the server sets positive contract items for type-1, type-2, and type-3 users  and zero contract item for other users, which validate Theorem \ref{thm4}. The threshold type users (i.e., type-3 users) only obtain a zero payoff, as the server's designed optimal rewards just cover their training costs and network costs. Type-1 and type-2 users (with marginal costs smaller than type-3 users) will obtain positive payoffs.%This validate the results in Theorem \ref{}

%\begin{figure}
%	\centering
%	\includegraphics[width=2.6 in]{cost1.eps}
%	\vspace{-3mm}
%	\caption{Network usage distribution and each user's network costs at different time slots under the vertical structure.}
%	\label{cost}
%\end{figure}

%\begin{figure}
%	\centering
%	\includegraphics[width=2.6 in]{server1.eps}
%	\vspace{-3mm}
%	\caption{Server's optimal contract and users' payoffs for different types of users under the vertical structure.}
%	\label{server}
%\end{figure}

\vspace{-1.5mm}
\subsection{Performance Comparison with Benchmarks}
\vspace{-1mm}
\label{ben}
To evaluate the performance, we list two benchmarks and our proposed mechanism as follows.
\vspace{-0.5mm}
\begin{itemize}
	\item \emph{No Joint Optimization (NJO) \cite{jiao2020toward}}: the server designs the incentive mechanism without  considering the network operator's strategies.
	\item \emph{No Dynamic Pricing (NDP)}: the server designs the incentive mechanism by assuming that the network operator sets a same price in all time slots\footnote{Due to   space limit, we will not present the closed-form optimal incentives and prices in	the benchmark cases.}.
	\item \emph{Our proposed pricing mechanism (IJD)}: the server designs the \underline{I}ncentive mechanism with the \underline{J}oint consideration of network operator's optimal \underline{D}ynamic pricing.
\end{itemize}
\vspace{-0.5mm}
%\begin{figure}
%	\centering
%	\includegraphics[width=2.6 in]{sercost.eps}
%	\vspace{-3mm}
%	\caption{The comparison of server's costs under three mechanisms.}
%	\label{sercost}
%\end{figure}
%\begin{figure}
%	\centering
%	\includegraphics[width=2.6 in]{propay.eps}
%	\vspace{-3mm}
%	\caption{The comparison of network operator's profits and users' total payoffs  under three mechanisms.}
%	\label{propay}
%\end{figure}

As shown in Fig.~\ref{sercost} and Fig.~\ref{propay}, our proposed mechanism outperforms the NJO and NDP benchmarks in terms of the server's cost, the network operator's profit, and users' total payoff. Compared with the NJO benchmark, the server's cost reduction and network operator's profit growth of our IJD mechanism reach 24.87\% and 1245.25\%, respectively.

%\begin{figure}[H]
%	\centering
%	\subfigure[Network usage and optimal prices.]{
%		\begin{minipage}[t]{0.48\linewidth}
%			\centering
%			\includegraphics[width=3 in]{price.eps}
%			%\caption{Coordinated interaction  structure.}
%			\label{coor}
%	\end{minipage}}
%	\subfigure[Network usage and users' network cost.]{
%		\begin{minipage}[t]{0.48\linewidth}
%			\centering
%			\includegraphics[width=3 in]{cost.eps}
%			%\caption{Horizontally-uncoordinated interaction  structure.}
%			\label{pull}
%	\end{minipage}}
%	\caption{Network usage, network operator's optimal prices, and each user's network costs at different time slots under the vertical structure.}
%	\label{structures}
%\end{figure}

%
%\subsection{Horizontal Structure}

%\begin{figure}
%	\centering
%	\includegraphics[width=2.8 in]{price2.eps}
%	\caption{Network usage and network operator's optimal prices at different time slots in both horizontal and  vertical structure.}
%	\label{price2}
%\end{figure}
%
%\begin{figure}
%	\centering
%	\includegraphics[width=2.8 in]{cost2.eps}
%	\caption{Network usage and each user's network costs at different time slots in both horizontal and  vertical structure.}
%	\label{cost2}
%\end{figure}
%
%
%\begin{figure}
%	\centering
%	\includegraphics[width=2.8 in]{server2.eps}
%	\caption{Server's optimal contract and users' payoffs for different types of users in both horizontal and  vertical structure.}
%	\label{server2}
%\end{figure}

\vspace{-1.5mm}
\section{Conclusion}
\label{conclusion}
\vspace{-1.5mm}
To the best of our knowledge, this is the first study on the important issue of joint participation incentive and network pricing design in federated learning. 
We   compared two typical interaction structures of the server,   network operator, and users in the system. We showed that the vertical interaction structure is better than the horizontal one for all participants. %, as it avoids the case where   the server  and the network operator target at different groups of users. 
Moreover, we demonstrated that  when users are congestion-sensitive, time slots with less background network demands encourage the federated learning users' selection  but still have less total network demands. 
The simulations showed that our proposed  mechanisms decrease the server's cost by up to 24.87\% and	increase network operator's profit  by up to 1245.25\%, compared with the state-of-the-art benchmarks. 
%For the future work,  we will consider the joint network pricing and incentive  design for competing network operators and  servers. %, when multiple servers are interested in using the data from the same pool of users to train similar types of machine learning models.
%the server wants users to upload results as soon as possible and he aggregates as soon as all users finish uploading. In that case, the training cycle may not be synchronized with the network pricing cycle, e.g., training cycle is shorter/longer than a day. The time of each training round/cycle can also change. In this case, the optimization in each training round will be different. Also, server's reward will be both time-dependent and data-dependent. %difference is adding one more dimension, time, in server's optimization problem, accuracy loss and reward. users' reward also different

%\vspace{-2.5mm}

\bibliographystyle{IEEEtran}
\bibliography{ref} 

\end{document}